\documentclass[aps,pra,
reprint,
superscriptaddress,
amsmath,amssymb,
floatfix,
]{revtex4-2}

\usepackage[caption=false,label font=bf]{subfig}
\usepackage{floatrow}
\usepackage{graphicx}
\usepackage{dcolumn}
\usepackage{bm}
\usepackage{braket}
\usepackage{hhline}
\usepackage{array}
\usepackage{booktabs}
\usepackage{nicefrac}
\usepackage[normalem]{ulem} 

\newcommand{\tens}[1]{%
  \mathbin{\mathop{\otimes}\displaylimits_{#1}}%
}
\usepackage[dvipsnames]{xcolor}
\newcommand{\avg}[1]{\left\langle #1 \right\rangle}

\definecolor{burgundy}{rgb}{0.5, 0.0, 0.13}

\newcommand{\revcom}[1]{{{#1}}} 

\usepackage{seqsplit}
\usepackage[hidelinks,colorlinks]{hyperref}
\hypersetup{citecolor=[rgb]{0.25,0.14,0.63}}
\hypersetup{urlcolor=[rgb]{0.25,0.14,0.63}}
\hypersetup{linkcolor=black}

\begin{document}

\preprint{APS/123-QED}

\title{A matter wave Rarity-Tapster interferometer to demonstrate non-locality}

\author{Kieran F. Thomas}
\affiliation{Department of Quantum Science and Technology, Research School of Physics, The Australian National University, Canberra, ACT 2601, Australia}
\author{Bryce M. Henson}

\author{Yu Wang}
\affiliation{Department of Quantum Science and Technology, Research School of Physics, The Australian National University, Canberra, ACT 2601, Australia}

\author{Robert J. Lewis-Swan}
\affiliation{Homer L. Dodge Department of Physics and Astronomy,
The University of Oklahoma, Norman, OK 73019, USA}
\affiliation{Center for Quantum Research and Technology, The University of Oklahoma, Norman, OK 73019, USA}

\author{Karen V. Kheruntsyan}
\affiliation{School of Mathematics and Physics, The University of Queensland, Brisbane, Queensland 4072, Australia}

\author{Sean S. Hodgman}
\affiliation{Department of Quantum Science and Technology, Research School of Physics, The Australian National University, Canberra, ACT 2601, Australia}
\author{Andrew G. Truscott}

\affiliation{Department of Quantum Science and Technology, Research School of Physics, The Australian National University, Canberra, ACT 2601, Australia}

\date{\today}

\begin{abstract}
We present an experimentally viable approach to demonstrating quantum non-locality in a matter wave system via a Rarity-Tapster interferometer using two \(s\)-wave scattering halos generated by colliding helium Bose-Einstein condensates. The theoretical basis for this method is discussed, and its suitability is experimentally quantified. As a proof of concept, we demonstrate an interferometric visibility of \(V=0.42(9)\), corresponding to a maximum CSHS-Bell parameter of \(S=1.1(1)\), for the  Clauser-Horne-Shimony-Holt (CHSH) version of the Bell inequality, between atoms separated by \(\sim 4\) correlation lengths. This constitutes a significant step towards a demonstration of a Bell inequality violation for motional degrees of freedom of massive particles and possible measurements of quantum effects in a gravitationally sensitive system.
\end{abstract}

\maketitle

\section{Introduction}


Quantum mechanics consistently defies our intuitive understanding of reality. It has introduced concepts such as wave-particle duality, where quantum objects can not be fully described using only a particle or wave representation, and entanglement. Together, these concepts have the remarkable implication that if quantum mechanics is complete, then nature itself appears to be inconsistent with descriptions based on the tenet of local realism \cite{RevModPhys.81.865}. The term local realism refers collectively to the principles of ``locality", which is that objects can only be influenced by their immediate surroundings, and ``realism", which asserts that an object's physical properties exist independently of a measurement by an observer. If one wishes to restore a notion of local realism one must assume quantum mechanics is incomplete and supplement it with a more fundamental theory. To this end, a number of local-hidden-variable interpretations of quantum mechanics have been proposed \cite{PhysRev.47.777,PhysRev.85.166,*PhysRev.85.180}. In these interpretations it was postulated that there was a more fundamental theory underpinning quantum mechanics, in which physical quantities were governed by well-defined (i.e., not probabilistic) yet inaccessible variables. In 1964  Bell proposed his famous Bell inequality, which is a set of conditions that all possible local-hidden-variable theories with freedom of choice must obey \cite{PhysicsPhysiqueFizika.1.195}, hence making the tenet of local realism experimentally testable. 

Over the subsequent decades numerous experimental investigations have been conducted, all of which have refuted local-hidden-variable theories and are consistent with the predictions of quantum mechanics \cite{PhysRevLett.28.938,PhysRevLett.47.460,PhysRevLett.49.91,PhysRevLett.49.1804,PhysRevLett.64.2495,PhysRevLett.81.3563,Hensen2015}. So far, experimental violations of Bell inequalities have been largely confined to the domain of quantum optics \cite{RevModPhys.86.419}. This is unsurprising, as photons can be manipulated and prepared in useful quantum states with relative ease. The comparative difficulty of preparing massive systems in a specific quantum state has led to a dearth of equivalent experimental investigations, particularly of Bell's inequality, in these systems \cite{RevModPhys.86.419}. While there have been demonstrations of violations of Bell's inequality using the internal states of atoms \cite{Rowe2001, PhysRevLett.91.110405, PhysRevLett.100.150404, doi:10.1126/science.1221856, Shin2019}, there has so far been no violation with massive particles using external degrees of freedom of those particles, such as momentum, i.e., degrees of freedom directly coupled to gravity. This is especially relevant as the incompatibility of quantum mechanics and general relativity is one of the deepest and most difficult problems of modern physics \cite{RevModPhys.61.561,rovelli2004quantum,Ashtekar_2004,kiefer2007quantum}. 


Our goal for the work presented here is to develop, demonstrate, and characterize a reliable platform for generating entanglement between external degrees of freedom of massive particles, specifically momentum states, and measuring phase-sensitive correlations between these entangled states. Such a platform would allow us to investigate a number of important research areas in the intersection of quantum mechanics and gravity. For instance, why are non-local correlations between massive objects in general uncommon and difficult to observe \cite{RevModPhys.81.865, RevModPhys.86.419}; how do we properly quantify entanglement in massive systems \cite{Zych_2011,Pikovski_2015,Kaltenbaek_2015}; and how does the size of a Bell violation scale with the size of the system \cite{LewisSwan2016,Penrose96}. Furthermore, pushing our understanding of quantum theory into realms which are increasingly more strongly coupled to gravity may enable the formulation and verification of a complete theory of quantum gravity \cite{Penrose96,Zych_2011,Pikovski_2015,PRXQuantum.2.030330,PhysRevLett.119.240402,PhysRevLett.119.240401}.

\begin{figure}[!htbp]
    \centering
    \includegraphics[width=0.735\textwidth]{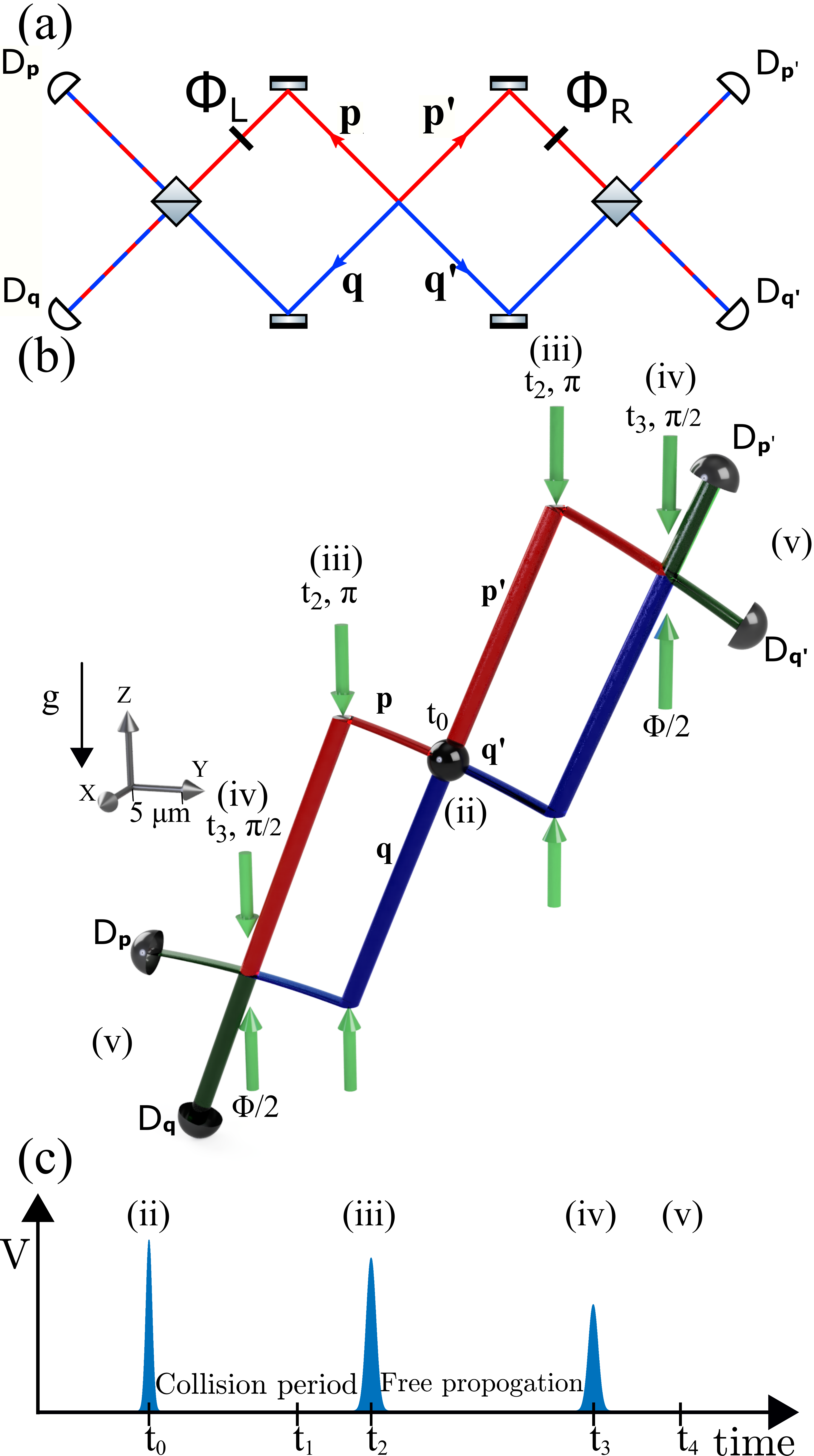}
    \caption{
    (a) Schematic of a modified optical Rarity-Tapster interferometer. Two pairs of momentum-entangled photons, labelled by $(\mathbf{p},\mathbf{p}')$ (red) and $(\mathbf{q},\mathbf{q}')$ (blue), are produced from a central source. The pairs are reflected onto one another such that they spatially overlap at a later beamsplitter. The indistinguishability of the pairs after the beamsplitter generates interference that depends on phase shifts $\phi_L$ and $\phi_R$ introduced in the upper arms of the interferometer.
    (b) Schematic of a matter wave Rarity-Tapster interferometer, indicating atom position space trajectories for a specific pair of modes which satisfy our interference conditions (the same modes highlighted in Fig.~\ref{fig:concept} and Roman numeral labels correspond to the same experimental stage). Trajectories are to scale for our parameters (with 5 $\mu$m scale indicated on the $y$-axis). The correlated pairs are produced by BEC collisions initiated at step (ii). The source BEC is transferred to the \(m_J=0\) state at step (i), which is not shown for clarity. At \(t_2\) we apply a \(\pi\) (mirror) Bragg pulse, step (iii), which reverses the atom's motion along $\hat{\textbf{z}}$. A phase difference (\(\phi_L=\phi_R=\Phi/2\)) between the input modes is imprinted via a \(\frac{\pi}{2}\) (beamsplitter) Bragg pulse at \(t_3\), step (iv). As the same set of lasers form the beamsplitter pulse for both arms, an equal phase is applied to both sides, and hence the net global phase shift is \(\phi_L+\phi_R=\Phi\).
    (c) Timeline of the experiment, vertical axis indicating intensity of Bragg light. Timings of steps (ii)-(v) are indicated, with step (i) not shown for clarity.
}
    \label{fig:concept_pos_sapce}
\end{figure}

\begin{figure*}
    \centering
    \includegraphics[width=0.95\textwidth]{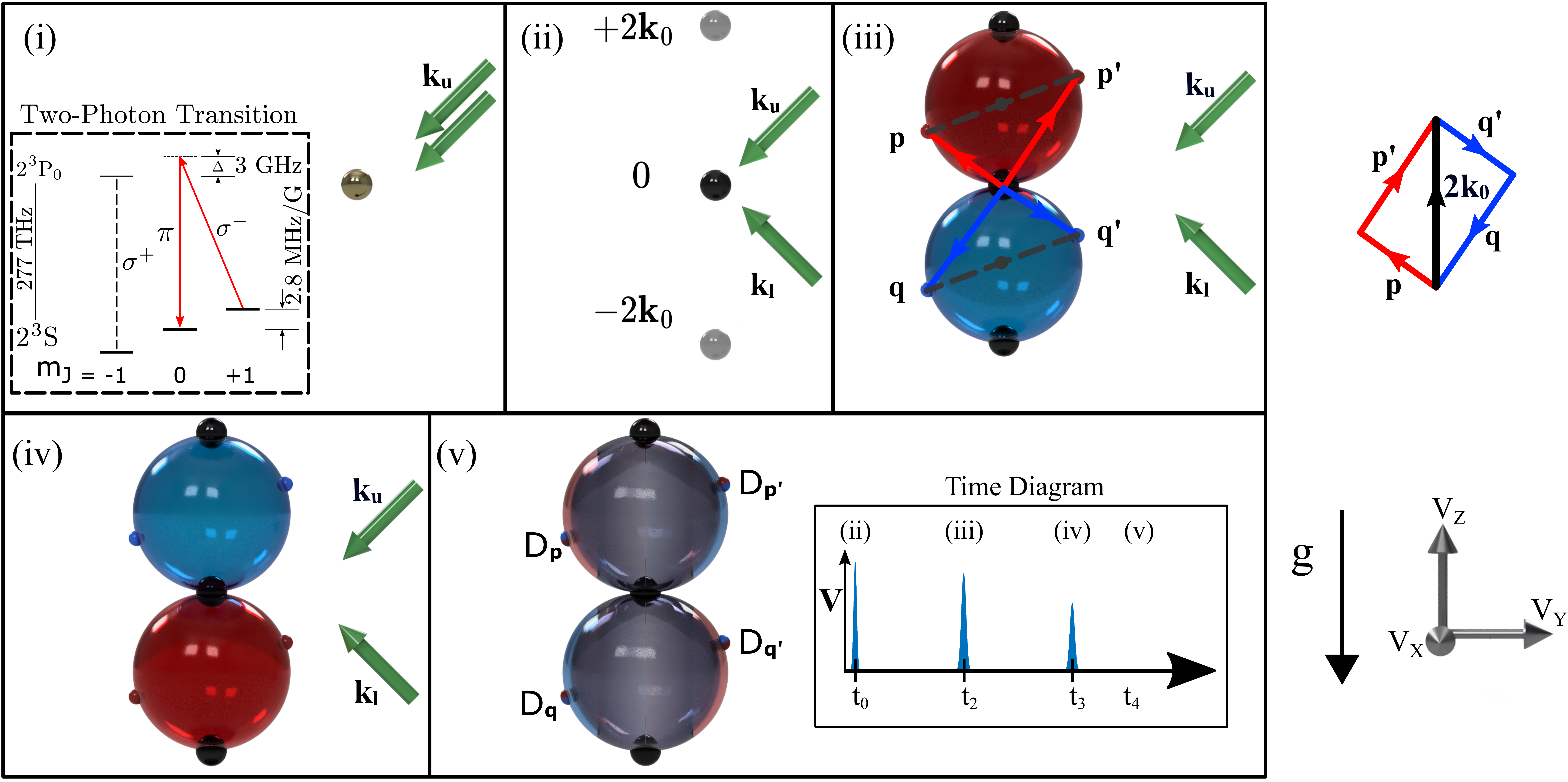}
    \caption{Schematic of the experimental procedure in momentum space. Gravity is aligned anti-parallel to the \(z\)-axis. Green arrows indicate laser beams used for the Bragg and Raman transitions, with \(\textbf{k}_u\) and \(\textbf{k}_l\) the wave vector of the upper and lower beams. (i) The initial He\(^*\) BEC is in the magnetically sensitive \(m_J=+1\) sublevel. A stimulated Raman transition, via a single beam with two frequency components, equivalent to two collinear lasers of differing frequency, is used to spin flip the atoms to the \(m_J=0\) state. (ii) Using two lasers intersecting at \(90^\circ\) (green arrows) we form a Bragg lattice (with Bragg vector \(\textbf{k}_0\) aligned parallel to the \(z\)-axis), which we use to split the BEC \revcom{equally} into momentum modes \(+2\textbf{k}_0,\, 0,\, -2\textbf{k}_0\). (iii) As the BEC components separate spatially, \(s\)-wave collisions between individual pairs of atoms from different BECs populate their corresponding scattering halos, highlighted in red for the scattering halo initially between \(+2\textbf{k}_0\) and \(0\) and blue for the halo between \(-2\textbf{k}_0\) and \(0\). Diametrically opposed atoms within each halo form a scattering pair due to momentum conservation, and hence are entangled.
    We mark an example set of \(4\) modes which satisfy our interference conditions (Eqs.~\eqref{eqn:int_conditions} and \eqref{eqn:int_conditions2}) with a dot and colour them to their respective initial halos to allow us to follow their path through momentum space more clearly. We also connect these points to the origin with coloured lines to highlight their connection to the position space version of the interferometer presented in Fig.~\ref{fig:concept_pos_sapce}, and highlight back to back pairs with gray dashed lines. Right of part (iii) we indicated how the pairs conserve the momentum of the collision. We apply a second Bragg pulse, which imparts an equal and opposite linear shift to the halos, hence acting as a mirror. This is necessary in order to overlap the atoms in position space at the mixing (beamsplitter) stage. (iv) We apply a beamsplitter Bragg pulse by imparting an equal and opposite shift to 50\% of the population, mixing the modes and making the interferometer outputs indistinguishable. (v) We label the respective detector outputs $D_{\textbf{k}}$ ($\textbf{k} \in \{ \mathbf{p},\mathbf{p}^{\prime} ,\mathbf{q},\mathbf{q}^{\prime}\}$) for the highlighted modes, in addition to showing a timing diagram of the voltage (\(V\)) sent to the acousto-optic modulators controlling our Bragg beams, with labels for the corresponding step of the experiment.
    }
    \label{fig:concept}
\end{figure*}

To develop this platform we take the path of extending the field of quantum optics to matter-waves. De Broglie postulated that all matter had an associated wavelength \cite{DeBroglie}, and thus could exhibit wave-like behaviour. With the advent of laser cooling techniques, the experimental generation of ultracold bosons and fermions became possible \cite{Phillips:85,Metcalf:03}. At these temperatures, the de Broglie wavelengths of the atoms are on the scale of tens or hundreds of micrometers, greatly increasing the scope of possible experimental investigations using matter waves. Creating atomic analogues of quantum optical systems have proved a valuable tool in pushing foundational tests of quantum mechanics into the realm of massive systems \cite{10.1038.416219a}, and has helped create the field of quantum atom optics \cite{Jeltes2007,Manning2015,Lopes2015,Manning:10,PhysRevLett.119.173202,Shin2019,PhysRevA.98.033608,PhysRevLett.126.083603}. These results point to cold atom optics as being a prime candidate system for an experimental violation of Bell's inequality with massive external degrees of freedom \cite{PhysicsPhysiqueFizika.1.195,stapp1975bell,Bell2004}.

In this paper, we present and quantify a method for conducting a Bell's inequality violation using pair-correlated scattered atoms from three colliding metastable helium (He\(^*\)) Bose-Einstein condensates (BECs). Our methodology is explained in Sec.~\ref{sec:background}, and builds on the ideas proposed in Refs. \cite{PhysRevA.91.052114,LewisSwan2016}. This represents an atomic realization of a Rarity-Tapster type interferometer \cite{PhysRevLett.64.2495} (see Fig.~\ref{fig:concept_pos_sapce}) and a continuation of our previous experiments using such pair-correlated atoms \cite{PhysRevLett.105.190402,PhysRevLett.108.260401,Khakimov2016,PhysRevLett.118.240402,Shin_2020}. 
We present our experimental evidence of spatial separated interference in Sec.~\ref{sec:results} and summarise the possibility of using this system to demonstrate a massive particle Bell violation and other further experiments in Sec.~\ref{sec:con}.



\section{Background}
\label{sec:background}
\subsection{Experimental implementation of a double-halo Rarity-Tapster interferometer}


The system we use in our experiment is an atomic BEC of helium atoms in the long lived \(2^3 S_1\) metastable state. We initially form our BEC in a biplanar quadrupole Ioffe magnetic trap \cite{Dall2007}, which is well approximated as harmonic near its minimum. For our experiment we relax our trap such that it has trapping frequencies \(\{\omega_x, \omega_y, \omega_z\}/2\pi \sim  \{50,200,200\}\)~Hz.
Approximately \(850\)~mm below the trap we employ a micro-channel plate (MCP) and delay line detector (DLD) system which allows us to detect the positions of single atom impacts after the atoms are released from the trap and allowed to fall in free space under gravity \cite{Manning:10}. Due to the distance the atoms travel, we can approximate these detection events as being in the far-field, enabling us to map the detected position of the atoms back to their initial velocity before expansion, with full 3D resolution. Thus, our system acts as a quantum many-body momentum microscope \cite{PhysRevLett.118.240402}, allowing us to sample the momentum distribution of the in-trap state and making this an ideal system to measure momentum correlations.



A schematic of our experimental producedure to implement a matter wave Rarity-Tapster interferometer \cite{PhysRevLett.64.2495} is shown in Figs.~\ref{fig:concept_pos_sapce} and \ref{fig:concept}. Our proposed setup is based on the original optical analog of the interferometer [Fig.~\ref{fig:concept_pos_sapce}(a)], which enables the measurement of phase-sensitive correlations between momentum-entangled twin-photons generated by a common source. This is achieved by mixing photons (or atoms) from independent pairs, labelled $(\mathbf{p},\mathbf{p}')$ and $(\mathbf{q},\mathbf{q}')$ in Fig.~\ref{fig:concept_pos_sapce}(a) (where we have used bold font to indicate vector quantities), at a pair of spatially separated beam-splitters after application of a pair of tunable phase-shifts in the upper arms of the interferometer. The trajectories through the interferometer to the detection ports $D_{\textbf{k}}$ ($\textbf{k} \in \{ \mathbf{p},\mathbf{p}^{\prime} ,\mathbf{q},\mathbf{q}^{\prime}\}$) are such that it is not possible to distinguish the original pairs, which enables detection of interference between the possible paths that is sensitive to the applied phase shifts. In the optical version the applied phases (\(\phi_L\) and \(\phi_R\)) is varied via optical elements inserted into the upper arms of the interferometer, in our case the phases will be operationally applied via the beamsplitter equivalent.

Our atomic analogue of this scheme is realized by applying a series of Raman and Bragg transitions, which utilise the \(2^3P_0\) state, after switching off the confining trap.
We first use a stimulated Raman transition, consisting of two collinear lasers of differing frequency, to transfer the BEC from the magnetically sensitive \(2^3 S_1 \, \, \, m_J = +1\) state to the insensitive \(2^3 S_1 \, \, \, m_J = 0\) state [Fig.~\ref{fig:concept}(i) inset] \cite{Shin2019}. As the transition uses collinear lasers no momentum is imparted \cite{Shin2019}. While the two transitions being stimulated are optimal for different polarisations we choose a single orientation that drives both transitions, i.e. is an equal combination of \(\pi\) and \(\sigma^-\) light, however, there is hence some power in the beam which is not used. To form our Bragg lattice, which allows us to control the momentum-state of the atoms (see Appendix~\ref{sec:bragg}), we use two \(276767\)~GHz approximately collimated Gaussian laser beams (whose wavevectors we label \(\textbf{k}_u\) and \(\textbf{k}_l\)) with \(1/e^2\) radius of \(0.7\)~mm intersecting at an angle of \(90^\circ\), with the resulting Bragg vector aligned along the \(z\)-axis (i.e., antiparallel to gravity). The spatial cross-section of the Bragg lasers is much larger than the atom paths (\(0.7\)~mm compared to a maximum atomic separation of \(\sim 60 \, \mu\)m) hence providing a uniform illumination across the entire halos and atom path. As these beams intersect at right angles, they generate a two-photon momentum recoil of \(2\hbar \textbf{k}_0 = \hbar(\textbf{k}_l-\textbf{k}_u) = \sqrt{2} \hbar k \hat{\textbf{z}}\), where \(k\) is the wavenumber of the incoming beams (\(|\textbf{k}_u|=|\textbf{k}_l|=k\)), and \(\hat{\textbf{z}}\) is a unit vector along the \(z\)-axis. For our chosen frequency of Bragg lasers, we thus have a Bragg vector of \(2\textbf{k}_0=8.204\, \hat{\textbf{z}}\)~\(\mu\)m\(^{-1}\). We then apply, at \(t_0\), a Bragg pulse that coherently splits the BEC into three momentum components, whose momenta have been respectively displaced by \(+2\textbf{k}_0,\, 0,\, -2\textbf{k}_0\) [see Fig.~\ref{fig:concept}(ii)]. \revcom{We tune the pulse such that the populations of the three momentum modes are approximately equal.}

As the moving condensates spatially separate, collisions between constituent atoms will occur. Provided that the BECs are in the slow moving regime these will be limited to \(s\)-wave, spherically symmetric elastic scattering events. From these, a pair of so-called \(s\)\textit{-wave collision halos} form in momentum space that, as a result of conservation of momentum and energy, are nearly spherical shells centered on the center-of-mass (COM) of each of the adjacent BEC pairs [Fig.~\ref{fig:concept}(iii) and \ref{fig:concept_pos_sapce}(b)]  \cite{PhysRevLett.104.150402,PhysRevLett.118.240402}. More precisely, the collision halos are composed of an ensemble of \revcom{independent two-mode-squeezed states, with each pair of squeezed modes} occupying diametrically opposed momenta due to conservation of momentum and energy, which can be used as the input state for our matter wave Rarity-Tapster interferometer.
Collisions between the \(+2\textbf{k}_0\) and \(-2\textbf{k}_0\) condensates also generate a larger halo, but this does not participate in the remainder of the dynamics and for clarity we omit it from Fig.~\ref{fig:concept}. We point out that our proposal to leverage twin-atoms generated in two collisonal halos is an important distinction from the previous theoretical proposal (based on a single halo) by two of the authors, Ref.~\cite{PhysRevA.91.052114}, and we elaborate on the consequences in Sec.~\ref{sec:Theory}.



After the halos are populated (we assume they are populated up to time \(t_1\)) we apply a series of Bragg pulses to couple a select set of momentum modes such that their paths through the interferometer are indistinguishable, and hence interfere. We refer to this setup as a Rarity-Tapster type matter wave interferometer \cite{PhysRevLett.64.2495} as it operates on the same principle as the optical version. Figure \ref{fig:concept}(iii-v) shows the momentum space picture and Fig.~\ref{fig:concept_pos_sapce} shows the trajectories through position space for this interferometer, with comparison to the optical counterpart. The timings of the pulses (mirror pulse at \(t_2\) and beamsplitter pulse at \(t_3\)), equivalent to the spatial positioning of the mirrors in the optical interferometer, must be optimised so that they provide maximum overlap between the scattering pairs, both in momentum and position space \cite{PhysRevA.91.052114}. 



\subsection{State of the scattering halos \label{sec:halo_state}} 

As previously mentioned, the collision halos are ideally composed of an ensemble of entangled twin-atoms with opposing momenta. Here, we elaborate on this statement by first considering the upper collision halo (formed by atoms scattering from the BECs with momenta $0$ and $+2\mathbf{k}_0$), before discussing how twin-atoms, scattered into the distinct upper and lower collision halos, approximately realize a prototypical Bell state that can be used to demonstrate violations of a Bell inequality. \revcom{Specifically, the mode-entanglement between squeezed modes will be turned into particle-like entanglement between a quartet (two pairs) of modes via post-selection, which is practically approximated using the low-gain regime.}


The collision of a pair of condensates realizes a matter-wave analog of four-wave mixing from quantum optics \cite{RuGway2011,Perrin2008}. Moreover, if the finite momentum width of the prepared condensates and depletion of its occupation due to the scattering of atoms into new momentum states is ignored, then we can approximately describe the resonant scattering of atoms from a pair of colliding condensates using a sum of two-mode squeezing Hamiltonians, \(\hat{H} = \sum_{\textbf{p}} \hbar \zeta \left( \hat{a}_{\textbf{p}}^{\dagger} \hat{a}_{\textbf{p}'}^{\dagger} +\hat{a}_{\textbf{p}'} \hat{a}_{\textbf{p}} \right)\) with $\textbf{p}'=-\textbf{p}+2\textbf{k}_0$. Here, \(\hat{a}_{\mathbf{p}(\mathbf{p}')}^{\dagger}\) creates a boson with momentum $\mathbf{p} \,(\mathbf{p}')$ and \(\zeta\) is an effective nonlinearity that is dependent on the \(s\)-wave scattering strength and 
the colliding BECs' densities \cite{PhysRevA.91.052114}, assumed to be uniform for simplicity.
Note that the momenta of the scattering pair must satisfy the condition \(\textbf{p}+\textbf{p}'=2\textbf{k}_0\), where \(2\textbf{k}_0\) is the sum of the colliding BECs' momenta, as we have assumed resonant scattering (e.g., energy is precisely conserved) and ignored the momentum width of the source BECs. Note, further that if we consider the system with respect to the COM frame of the halo, then our description of the scattered pairs reduces to that of previous works \cite{PhysRevA.91.052114,LewisSwan2016}.

Considering now the collision of $0$ and $+2\mathbf{k}_0$ BECs and assuming an initial vacuum state for all momenta $(\mathbf{p}, \mathbf{p}') \neq 0$ or $2\mathbf{k}_0$, the two-mode squeezing Hamiltonian generates a product of two-mode squeezed vacuum states for the upper halo, which can be expressed in the Fock basis as \cite{SQUEEZING_NOTE,RevModPhys.77.513},
\begin{align}
    \ket{\psi}_{\mathrm{halo,upper}} &= \prod_{\mathbf{p}}\otimes\left( \sqrt{1-\mu^2} \sum^{\infty}_{n=0} \mu^{n} \ket{n}_{\mathbf{p}}\ket{n}_{\mathbf{p}'} \right), \label{eqn:state}
\end{align}
where $\mathbf{p}'=-\mathbf{p}+2\mathbf{k}_0$, \(\mu=\tanh(\zeta t_1)\) for a collision duration $t_1$, and $\zeta t_1$ is the effective squeezing parameter. Thus, Eq.~\eqref{eqn:state} represents a product state composed of superpositions of number-correlated states $\ket{n}_{\mathbf{p}}\ket{n}_{\mathbf{p}'}$ (in any pair of modes with momenta $\mathbf{p}$ and $\mathbf{p}'$  satisfying $\mathbf{p}+\mathbf{p}'=2\mathbf{k}_0$) with an increasing $n$ and a decreasing probability amplitude $\propto \!\mu^{n}$, where $\mu<1$. Further, if we consider a single mode, its occupation is described by thermal counting statistics \cite{10.21468/SciPostPhys.7.1.002}.
The mean number of atoms in a particular scattering mode is given by \(\bar{N}=\frac{\mu^2}{1- \mu^2}=\sinh^2(\zeta t_1)\). Therefore,  \(\bar{N} \approx \mu^2\ll 1\) for \(\mu \ll 1\), i.e., in the low-gain or perturbative regime that we are going to consider below. 


Under the same  assumption as before---that the depletion of colliding condensates is negligible and the initial state of the halo is a vacuum state---the state of the second (lower) halo, which is formed by the collision between the \(-2\textbf{k}_0\) and \(0\) BECs, is independent of the upper halo. Therefore, the lower halo can be identically described by Eq.~\eqref{eqn:state}, albeit with corresponding $\mu' = \tanh(\zeta' t_1)$, 
\begin{align}
     \ket{\psi}_{\mathrm{halo,lower}} &= \prod_{\mathbf{q}}\otimes\left( \sqrt{1-\mu'^{2}} \sum^{\infty}_{n=0} \mu'^{n} \ket{n}_{\mathbf{q}}\ket{n}_{\mathbf{q}'} \right), \label{eqn:state_lower}
\end{align}
where \(\textbf{q}'=-\textbf{q}-2\textbf{k}_0\). We introduce the distinct \(\mu'\) to incorporate the possibility that, e.g., the density of the split BECs is not identical (leading to a different nonlinearity $\zeta'$) but assume the collision duration $t_1$ is unchanged. 
If all three colliding BECs are identical then \(\mu=\mu'\), which we will assume herein. The independence of the formed collision halos implies that 
the complete state describing all the (resonantly) scattered atom pairs in both halos is 
\begin{equation}\ket{\Psi}_{\text{double-halo}}= \ket{\psi}_{\mathrm{halo,upper}} \tens{} \ket{\psi}_{\mathrm{halo,lower}}.
\end{equation}

\subsection{Mapping of scattering state to a Bell state\label{sec:Theory}}

Now we will show how the input state of our interferometer, a reduced form of \(\ket{\Psi}_{\text{double-halo}}\), can be approximated to one of the prototypical Bell states. To do so, we focus on just two correlated pairs of momentum modes $(\mathbf{p},\mathbf{p}^{\prime})$ and $(\mathbf{q},\mathbf{q}^{\prime})$ selected from the upper and lower halos, respectively, such that they satisfy,
\begin{align}
   \textbf{p}+\textbf{p}'=2\textbf{k}_0,\, \quad \textbf{q}'+\textbf{q}=-2\textbf{k}_0,\label{eqn:int_conditions}
\end{align}
and simultaneously are related by,
\begin{align}
   \textbf{p} = \textbf{q} + 2\textbf{k}_0, \, \quad \textbf{p}' = \textbf{q}' + 2\textbf{k}_0.\label{eqn:int_conditions2}
\end{align}
We illustrate an example of the selected modes in Figs.~\ref{fig:concept_pos_sapce} and \ref{fig:concept}. Physically, the modes correspond to two scattering pairs \((\textbf{p},\textbf{p}')\) and \((\textbf{q},\textbf{q}')\), which are exactly separated by the two-photon momentum recoil \(2\textbf{k}_0\) used in the initial momentum slitting of the source BEC [see Fig.~\ref{fig:concept}(ii)]. They were chosen as they are the only modes, within the halos, which exactly match the Bragg condition, i.e., are separated by an integer number of lattice wave vectors, for the geometry used to generate the original BEC momentum splitting. This allows for a more stable and simple implementation, as discussed in further detail below. 

Ignoring or tracing away all other momentum mode pairs
from $|\Psi\rangle_{\mathrm{double-halo}}$ beyond the two pairs considered, the \revcom{factorised} state of the four modes of interest can be written as
\begin{align}
    \ket{\Psi} &\equiv \ket{\Psi}_{\mathbf{p},\mathbf{p}',\mathbf{q},\mathbf{q}'}\nonumber \\  &=(1-\mu^2) \sum^{\infty}_{n,m=0} \mu^{(n+m)} 
     \ket{n}_{\mathbf{p}}\ket{n}_{\mathbf{p'}}\ket{m}_{\mathbf{q}}\ket{m}_{\mathbf{q}'}. \label{eqn:reduced_state}
\end{align}
Assuming now that the scattering from the condensates is in the low-gain, perturbative  regime, \(\mu\ll1\), we can truncate the sum over Fock states to lowest order in $\mu$, equivalent to ignoring the contribution of Fock states with \revcom{the total occupancy among the four modes larger than 2 (i.e., restricting ourselves, via post-selection, to no more than a total of 2 particles across the two halos)}
\revcom{The removal of higher occupancy terms has the effect of entangling the pair of selected squeezed modes $(\mathbf{p},\mathbf{p}^{\prime})$ with the other selected pair $(\mathbf{q},\mathbf{q}^{\prime})$ , which previously had been independent of each other (non-entangled). It will be seen that this is why in the low-gain limit ($\mu \rightarrow 0$) the full state, despite being factorisable, is able to approximate a maximally particle entangled Bell state.} We enforce the low-gain, perturbative regime used in this analysis, and the experiment, by ensuring that $\zeta t_1 \ll 1$, which typically corresponds to either a short effective collision duration (due to, e.g., rapid spatial separation of the BECs) or low initial density of the source condensate. The resulting truncated state for the two pairs of correlated modes can therefore be written as

\begin{align}
    \ket{\Psi} &\approx (1-\mu^2) \left(\ket{0}_{\mathbf{p}}\ket{0}_{\mathbf{p'}}\ket{0}_{\mathbf{q}}\ket{0}_{\mathbf{q}'}\right.\nonumber\\
    &\left.+\mu \ket{0}_{\mathbf{p}}\ket{0}_{\mathbf{p'}}\ket{1}_{\mathbf{q}}\ket{1}_{\mathbf{q}'}+\mu \ket{1}_{\mathbf{p}}\ket{1}_{\mathbf{p'}}\ket{0}_{\mathbf{q}}\ket{0}_{\mathbf{q}'}\right),
    \label{eqn:truncated-state}
\end{align}

As the vacuum state $\ket{0}_{\mathbf{p}}\ket{0}_{\mathbf{p'}}\ket{0}_{\mathbf{q}}\ket{0}_{\mathbf{q}'}$, the first term in Eq.~\eqref{eqn:truncated-state}, does not contribute to any measurement of population correlations that we will consider below, we can further truncate the above expression to 
\begin{align}    
  \ket{\Psi}\approx \frac{1}{\sqrt{2}} \left(\ket{0}_{\mathbf{p}}\ket{0}_{\mathbf{p'}}\ket{1}_{\mathbf{q}}\ket{1}_{\mathbf{q}'}+ \ket{1}_{\mathbf{p}}\ket{1}_{\mathbf{p'}}\ket{0}_{\mathbf{q}}\ket{0}_{\mathbf{q}'}\right), \label{eqn:proto_bell}
\end{align}
where we have enforced normalization. 

This reduced and truncated state, describing the atom pairs in modes $(\mathbf{p},\mathbf{p}^{\prime})$ and $(\mathbf{q},\mathbf{q}^{\prime})$ in the low-gain regime, is formally equivalent to a paradigmatic polarization-entangled Bell state of two photons exploited in optical tests of quantum nonlocality. To make this clear we consider a mapping of atomic mode occupations in the left (L) and right (R) arms of the Rarity-Tapster interferometer [see Figs.~\ref{fig:concept_pos_sapce}\,(a-b) and \ref{fig:concept}\,(iii)] onto two, horizontal (H) and vertical (V), polarisation states of photons, 
\begin{align}
|0\rangle_{\mathbf{p}}|1\rangle_{\mathbf{q}} &\rightarrow |0\rangle_{\mathrm{V}}^{(\mathrm{L})}|1\rangle_{\mathrm{H}}^{(\mathrm{L})} \equiv |\mathrm{H}\rangle^{(\mathrm{L})},\\
|1\rangle_{\mathbf{p}}|0\rangle_{\mathbf{q}} &\rightarrow |1\rangle_{\mathrm{V}}^{(\mathrm{L})}|0\rangle_{\mathrm{H}}^{(\mathrm{L})}\equiv |\mathrm{V}\rangle^{(\mathrm{L})},
\end{align}
 corresponding to an atom in the left arm populating either the $\mathbf{q}$ or $\mathbf{p}$ momentum mode. Similarly, for the right arm, 
\begin{align}
|0\rangle_{\mathbf{p}'}|1\rangle_{\mathbf{q}'} &\rightarrow |0\rangle_{\mathrm{V}}^{(\mathrm{R})}|1\rangle_{\mathrm{H}}^{(\mathrm{R})} \equiv |\mathrm{H}\rangle^{(\mathrm{R})},\\
|1\rangle_{\mathbf{p}'}|0\rangle_{\mathbf{q}'} &\rightarrow |1\rangle_{\mathrm{V}}^{(\mathrm{R})}|0\rangle_{\mathrm{H}}^{(\mathrm{R})} \equiv |\mathrm{V}\rangle^{(\mathrm{R})},
\end{align}
corresponding to an atom in the right arm populating either the $\mathbf{q}'$ or $\mathbf{p}'$ momentum mode. Then, Eq.~\eqref{eqn:proto_bell} can be mapped to $\ket{\Psi} \rightarrow |\Psi\rangle_{\mathrm{Bell}}=\frac{1}{\sqrt{2}}(|H\rangle^{(\mathrm{L})} |H\rangle^{(\mathrm{R})} + |V\rangle^{(\mathrm{L})} |V\rangle^{(\mathrm{R})})$, or 
 \begin{equation}\label{eqn:BellstateHV}
 |\Psi\rangle_{\mathrm{Bell}}= \frac{1}{\sqrt{2}}(|H,H\rangle + |V,V\rangle).
 \end{equation} 
This can now be readily identified as the prototypical Bell state, which is known to maximally violate a Bell inequality \cite{PhysRevLett.49.1804,Aspect2004,nielsen00}. While for our purposes we will only consider pairs selected from different halos, any two \emph{independent} pairs of correlated modes, across either halo, can be mapped to one of the Bell states \revcom{\emph{after} post-selection}. Which Bell state exactly depends on the interferometeric geometry used; for example, the state  under the previously proposed scheme of Ref.  \cite{PhysRevA.91.052114}  will map to $\frac{1}{\sqrt{2}}(\ket{H,V}+\ket{V,H})$.

\revcom{We again wish to highlight} that the entanglement in our reduced state  $\ket{\Psi}$, Eq.~\eqref{eqn:reduced_state}, is identifiable as mode-entanglement, which is between back-to-back (diametrically opposed) momentum modes in a single halo. Only after we \revcom{remove} the contribution of states with single mode occupation higher than or equal to \(2\), which is equivalent in a physical sense to post-selecting for states that only contain a single pair of atoms among all four considered modes, can the state be reinterpreted as featuring particle Bell-state entanglement. In practice, robust post-selection is not possible due to the low detection efficiency that can be achieved in our experimental setup. Thus, our atomic four-wave mixing source acts as a \emph{probabilistic} generator of particle entangled Bell states, rather than a deterministic one as is usually considered. Hence, the need for many experimental runs in the low mode occupation regime; we require that in the relevant momentum modes the probability of mode occupancies of $2$ or higher is negligible, while retaining a small probability of desired occupancies of $1$ in a fraction of experimental runs.


\subsection{Theoretical basis of the matter-wave Rarity-Tapster interferometer}


The working principle of the matter wave Rarity-Tapster interferometer in our realization is as follows. The pairs of entangled atoms with momenta $(\mathbf{p},\mathbf{p}^{\prime})$ and $(\mathbf{q},\mathbf{q}^{\prime})$ are reflected onto spatially separated atomic beamsplitters with phases $\phi_L$ and $\phi_R$. 
The atomic beamsplitters cause the initially distinct entangled pairs to become indistinguishable and leads to multi-particle interference effects that are sensitive to the applied phases (see Fig.~\ref{fig:concept_pos_sapce}). 
We note that to exploit this multi-particle interference for a proper demonstration of quantum nonlocality we in principle require independent tunability of the phases $\phi_L$ and $\phi_R$ (see Sec.~\ref{sec:ctsmodel}). However, in our current realization we realize the atomic beamsplitters with a single Bragg laser, such that $\phi_L = \phi_R$, which is sufficient for us to benchmark the interferometer.
Our matter wave interferometer is equivalent in spirit to that employed by Rarity and Tapster in their experimental violation of the Bell inequality using the phase and momentum of photons \cite{PhysRevA.41.5139,PhysRevLett.64.2495}, but with a swapping of particle labels in one arm of the interferometer.

At the output of the interferometer (i.e., at some time $t_4$ after the final beam-splitter) we measure population correlations between the two pairs of momenta $(\mathbf{p},\mathbf{p}^{\prime})$ and $(\mathbf{q},\mathbf{q}^{\prime})$. To be concrete, we introduce the pair correlation function $C_{\mathbf{k},\mathbf{k}^{\prime}} = \langle \hat{a}^{\dagger}_{\mathbf{k}} \hat{a}^{\dagger}_{\mathbf{k}^{\prime}} \hat{a}_{\mathbf{k}^{\prime}}\hat{a}_{\mathbf{k}} \rangle$ (where \(\textbf{k}\in\{\textbf{p},\textbf{q}\}\) and \(\textbf{k}'\in\{\textbf{p}',\textbf{q}'\}\)), which corresponds to a correlation between joint-detection events at the outputs of the left and right arms of the matter wave interferometer (see Figs.~\ref{fig:concept_pos_sapce} and \ref{fig:concept}). In simple terms the two point correlation function gives us a measure of the probability of finding a similar number of particles at momenta \(\textbf{k}\) and \(\textbf{k}'\). It hence contains information about the spatially separate interference and will serve as the basis of our test of Bell's inequality. 

Treating the atomic mirror and beamsplitter elements of the interferometer as instantaneous linear transformations (see Appendix~\ref{sec:mapping} for details of the calculation), and using the reduced state, Eq.~\eqref{eqn:reduced_state}, as input, we obtain the normalized pair-correlations   
\begin{align}\label{eqn:C_kkp_tms}
    \frac{C_{\textbf{p}, \textbf{p}'}}{\bar{N}^2} &= \frac{C_{\textbf{q}, \textbf{q}'}}{\bar{N}^2} = 1 + \mu^{-2} \sin^2\left(\frac{\phi_L + \phi_R}{2}\right) , \\
    \label{eqn:theotherone}
    \frac{C_{\textbf{p}, \textbf{q}'}}{\bar{N}^2} &= \frac{C_{\textbf{p}', \textbf{q}}}{\bar{N}^2} = 1 + \mu^{-2} \cos^2\left(\frac{\phi_L + \phi_R}{2}\right), 
    \end{align}
where \(\bar{N}=\frac{\mu^2}{1- \mu^2}\) is the average occupancy of any of the four modes, $\mathbf{p}$, $\mathbf{p}^{\prime}$, $\mathbf{q}$, and $\mathbf{q}^{\prime}$. Here, we emphasize that interference between the scattered pairs, as captured by the sine and cosine terms in the above pair-correlation functions, is tuned via the \emph{global} phase \(\Phi=\phi_L+\phi_R\), rather than the \emph{relative} value of $\phi_L-\phi_R$.

In the regime $\mu \ll 1$ (i.e., vanishing probability of multiply occupied modes, $\bar{N} \ll 1$) the phase-sensitive terms $\propto \mu^{-2}$ dominate the correlation functions. However, when the mode occupancies  with 2 or more atoms  become appreciable, corresponding to $\bar{N} > 1$ (\(\mu \lesssim 1\)), we find that the oscillatory terms depending on the global phase $\phi_L + \phi_R$ diminishes relative to the background and thus interference is suppressed. 
Nevertheless, the undepleted pump approximation, used to derive Eq.~\eqref{eqn:reduced_state} and which ensures a highly nonclassical state of the scattered pairs, will eventually fail in the high-gain, nonperturbative limit.

We will find that this demonstrates that the matter wave Rarity-Tapster interferometer optimally probes quantum nonlocality when the initial state of the scattering halo can be closely \revcom{approximated} to a prototypical Bell state, i.e., when $\mu\ll 1$.

\revcom{It is interesting to note that, as we do not experimentally post-select on our final state, the modes $(\mathbf{p},\mathbf{p}^{\prime})$ by definition remain unentangled with $(\mathbf{q},\mathbf{q}^{\prime})$ as the state is factorisable. Specifically, there is no entanglement between states in different halos. Nonetheless, it can be seen from Eqs.~\ref{eqn:C_kkp_tms} and \ref{eqn:theotherone} that the mode-entanglement between $\mathbf{p}$ and $\mathbf{p}^{\prime}$, as well as between $\mathbf{q}$ and $\mathbf{q}^{\prime}$, in the initial state (i.e. between the left and right arms of the interferometer) is sufficient to violate Bell’s inequality for sufficiently small \(\mu\).}

\subsection{Continuous model for correlation functions \label{sec:ctsmodel}}

Experimentally, the finite momentum and spatial width of the initial BECs contributes to a finite correlation width of the scattered atoms, which may no longer be scattered with precisely correlated momenta as in Eq.~\eqref{eqn:int_conditions}. To account for this, we first introduce the integrated pair-correlation function between detection regions centered around \(\textbf{L}\in\{\textbf{p},\textbf{q}\}\) and \(\textbf{R}\in\{\textbf{p}',\textbf{q}'\}\) (see Figs.~\ref{fig:concept_pos_sapce} and \ref{fig:concept} for definition of port labeling), which is defined as,
\begin{align}
    \mathcal{C}_{\textbf{L},\textbf{R}} &=  \int_{V(\textbf{L})} d \textbf{k} \int_{V(\textbf{R})} d \textbf{k}'  ~ G^{(2)}(\textbf{k},\textbf{k}',t_4), \label{eqn:int_pair_corr}
\end{align}
where 
\(G^{(2)}(\textbf{k},\textbf{k}',t_4) \!=\! \avg{\hat{a}^{\dagger}(\textbf{k},t_4) \hat{a}^{\dagger}(\textbf{k}',t_4) \hat{a}(\textbf{k}',t_4) \hat{a}(\textbf{k},t_4) }\)
is the two-point momentum correlation function evaluated at time \(t_4\) after the interferometric sequence, \(V(\textbf{L})\) and \(V(\textbf{R})\) are the volumes of the respective detection regions in momentum space, whereas $\hat{a}^{\dagger}(\textbf{k},t_4)$ and $\hat{a}(\textbf{k},t_4)$ are the Fourier components of the field creation and annihilation operators describing the scattered atoms, evaluated at time $t_4$ in the Heisenberg picture \cite{PhysRevA.91.052114}. Note that \(\mathcal{C}_{\textbf{L},\textbf{R}}\) can be considered a generalization of \(C_{\mathbf{k},\mathbf{k}^{\prime}}\) where momenta are no longer treated as discrete.

The evaluation of \(\mathcal{C}_{\textbf{L},\textbf{R}}\) requires us to obtain a specific form for \(G^{(2)}(\textbf{k},\textbf{k}',t_4)\) for cases where \(\textbf{k} \approx \textbf{k}'\pm 2\textbf{k}_0\), and the sign depends on the relevant pair of detection regions under consideration. To do so, we again treat the Bragg pulses as instantaneous (and perfect) linear transformations. Moreover, we adopt a Gaussian approximation for the two-point correlation \(G^{(2)}(\textbf{k},\textbf{k}',t_1)\) of the input state \cite{PhysRevA.91.052114,PhysRevLett.100.233601,PhysRevLett.99.150405} between pairs of momenta $(\mathbf{k},\mathbf{k}^{\prime})$ within the \emph{same} scattering halo,
\begin{equation}
\frac{G^{(2)}(\textbf{k},\textbf{k}',t_1)}{n_0^2} =  1 + h  \prod_d \exp \left[\frac{-(k_d+k_d' \pm 2k_0\delta_{dz})^2}{2\sigma_d^2} \right] .\label{eqn:gaussian_g2}
\end{equation}
Here, \(h\) is the height of the correlation amplitude above the background level (generally assumed to be \(1\)), \(\sigma_d\) is the momentum correlation width in dimension \(d\) ($d=x,y,z$) of the original state, \(\delta_{ij}\) is the Kronecker delta function, and \(n_0\) is the peak momentum space density of scattered atoms. Note, if we take the two-mode squeezed state to characterize the halo, the correlation height \(h\) is related to the mode occupancy as \(h = 1/\mu^2 \approx 1/\bar{N}\), and \(n_0 \prod_d \sigma_d = \bar{N}\). The plus and minus signs in Eq.~\eqref{eqn:gaussian_g2} correspond to correlations between modes in the upper and lower halos, respectively. For pairs of momenta $(\mathbf{k},\mathbf{k}^{\prime})$ selected from distinct halos we have that $G^{(2)}(\textbf{k},\textbf{k}',t_1)/n_0^2 = 1$, as the scattered halos are assumed to be independent and uncorrelated.

Putting these pieces together, we obtain the relevant two-point correlation functions at the conclusion of the matter wave interferometer,
\begin{widetext}
\begin{equation}
\frac{G^{(2)}(\textbf{k},\textbf{k}',t_4)}{n_0^2} = 
\begin{cases}
1 + \frac{h}{2} \left[1 -  \cos \left( \Phi + \varphi(\textbf{k},\textbf{k}') \right) \right]  \prod_d \exp \left[\frac{-(k_d+k_d' \pm 2k_0\delta_{dz})^2}{2\sigma_d^2} \right], & \text{for} ~(\mathbf{k},\mathbf{k}') \approx (\mathbf{p},\mathbf{p}') ~ \text{or} ~ (\mathbf{q},\mathbf{q}'), \\[14pt]
1 + \frac{h}{2} \left[1 +  \cos \left( \Phi + \varphi(\textbf{k},\textbf{k}') \right) \right]  \prod_d \exp \left[\frac{-(k_d+k_d')^2}{2\sigma_d^2} \right], & \text{for} ~(\mathbf{k},\mathbf{k}') \approx (\mathbf{q},\mathbf{p}') ~ \text{or} ~ (\mathbf{p},\mathbf{q}'), 
\end{cases}\label{eqn:output_g2_full}
\end{equation}
\end{widetext}
where the sign of the cosine function depends on the particular detection port combination. 
Again, the sign of the term in the Gaussian depends on whether the relevant momenta are in the upper ($-$) or lower ($+$) halos. In addition to the Gaussian dependence on the correlation lengths that is inherited from $G^{(2)}(\textbf{k},\textbf{k}',t_1)$, a momentum-dependent phase offset $\varphi(\textbf{k},\textbf{k}')$ (see Appendix \ref{sec:phase_diffusion}) between off-resonant scattered pairs is included. We postpone a discussion of the physical meaning of this term to below.

The relevant integrated pair-correlation function, Eq.~\eqref{eqn:int_pair_corr}, is evaluated by substituting Eq.~\eqref{eqn:output_g2_full} and integrating over a detection bin with dimensions $\Delta k_d$ for $d = x,y,z$, leading to the result
\begin{align}
    \frac{\mathcal{C}_{\textbf{L},\textbf{R}}}{\bar{N}_V^2} &= 1 + \frac{h}{16} \left[1 \pm  \cos \left( \Phi \right) \right] \alpha_x \alpha_y \beta_z \prod_{d} \lambda_d^{-2} . \label{eqn:corr_final}
\end{align}
We have introduced the average number of particles in the integration volume, \(\bar{N}_V= n_0 \prod_{d} \Delta k_d\), and  \(\lambda_d \equiv \Delta k_d/(2 \sigma_d)\) as the normalized resolution of the detection bin 
(see Tab.~\ref{tab:params} for experimental values). Compared to the simpler form of Eq.~\eqref{eqn:C_kkp_tms}, the function \(\alpha_d\) for $d = x,y$ takes into account the contribution of the finite correlation widths in Eq.~\eqref{eqn:output_g2_full}, while the function \(\beta_d\) for $d = z$ also includes an additional contribution from the phase factor \(\varphi(\textbf{k},\textbf{k}')\) (which is in fact only a function of $k_z$ and $k_z^{\prime}$ for our interferometer configuration). These functions are defined as follows,

\begin{align}
    \alpha_d \equiv (e^{-2 \lambda_d^2}-1)+ \sqrt{2 \pi} \lambda_d \text{erf} \left( \sqrt{2} \lambda_d \right),
\end{align}
and,
\begin{align}
    \beta_d &\equiv e^{-2 \lambda_d^2} \cos \left( 4 A_d \lambda_d \right)-1+ 2\sqrt{2} A_d D_F \left(\sqrt{2} A_d\right)  \\
    &+ \sqrt{\frac{\pi}{2}} e^{-2 A_d^2} \Big[ (\lambda_d-iA_d) \text{erf} \left( \sqrt{2} \lambda_d - \sqrt{2} i A_d  \right) +\nonumber\\
    &\, \, \, \, \, (\lambda_d+iA_d) \text{erf} \left( \sqrt{2} \lambda_d + \sqrt{2} i A_d  \right) \Big],\nonumber
\end{align}
where \(A_d=\frac{k_0 \sigma_d \hbar}{m}(\frac{t_3}{2} - t_1)\) and 
\(D_F(x)=e^{-x^2}\int_0^x e^{y^2} dy\) is the Dawson \(F\) function. 

The dependence of $\mathcal{C}_{\textbf{L},\textbf{R}}$ on the size of the chosen detection bins captures the fact that for $\lambda_d \gg 1$ there are many uncorrelated atoms within each integration volume, and so we expect $\mathcal{C}_{\textbf{L},\textbf{R}}/\bar{N}_V^2 \to 1$. However, the integrated pair-correlation also encodes a subtle dependence on the relative timing of the mirror and beam-splitter Bragg pulses via the $\beta_z$ factor, and in turn the momentum dependent phase offset \(\varphi(\textbf{k},\textbf{k}')\) in Eq.~\eqref{eqn:output_g2_full}, from which it arises. This phase offset captures the importance of indistinguishability in the atomic realization of the Rarity-Tapster interferometer, in close analogy to work studying Hong-Ou-Mandel interference with atom pairs \cite{Lopes2015}. In particular, we see that $A_z \to 0$, and consequently $\beta_z \to \alpha_z$, by choosing $t_3 = 2t_2$, i.e., by setting the free propagation period $t_3 - t_2$ between the mirror pulse ($t_2$) and final beamsplitter pulse ($t_3$) to be identical to the prior time period between the commencement of the BEC collision and the application of the mirror pulse. This result is distinct to, and in fact much simpler than, that found in the prior study of Ref.~\cite{PhysRevA.91.052114}, which considered correlated pairs of momentum modes drawn from a single scattering halo. In that case, the optimal free propagation period $t_3 - t_2 = t_2 - m \sigma_{g,z}/k_0\sqrt{\pi}$ was offset by a factor equivalent to the effective collision duration set by the time taken for the colliding BECs to spatially separate, $t_{\mathrm{sep}} = \sigma_{g,z} m/\hbar k_0$ where $\sigma_{g,z}$ is the estimated rms spatial width of the initial unsplit condensate. This distinction between our scheme exploiting two scattering halos and that of Ref.~\cite{PhysRevA.91.052114} arises due to a subtle difference in the relative position along the splitting direction ($z$-axis) at which the scattered pairs are considered to be created as a function of collision duration. 

In our scheme, pairs located on, e.g., the equatorial $k_x-k_y$ plane of the scattering halo are always created at the COM of the colliding BEC pair, with each scattered pair also having COM momenta $\pm k_0\hat{\textbf{z}}$. Thus, if we consider the creation of scattered pairs as a classical stochastic process over time, all the created pairs have an identical COM position along the $z$-axis and travel with the same velocity along the direction of the applied Bragg pulses. This means that for $t_3 = 2t_2$ all scattered pairs share approximately the same $z$ position (up to, e.g., corrections due to the initial spatial width of the source condensates) when they are subject to the final beamsplitter Bragg pulse. Importantly, this implies that it is not possible to distinguish which pair of the originally coupled momentum modes, $(\mathbf{p},\mathbf{p}^{\prime})$ and $(\mathbf{q},\mathbf{q}^{\prime})$, an atom belongs to when it is finally detected and the integrated pair-correlations $\mathcal{C}_{\textbf{L},\textbf{R}}$ are constructed. This indistinguishability between the possible paths in the interferometer is pivotal to obtain phase-sensitive correlations, which is immediately seen by noting that $\mathcal{C}_{\textbf{L},\textbf{R}}/\bar{N}_V^2 \to 1$ for $A_d \gg 1$ (corresponding to a large mismatch between the free propagation periods before and after the mirror Bragg pulse).

In contrast, in Ref.~\cite{PhysRevA.91.052114} scattered pairs are always created approximately at the origin but at random times $0 \leq t \lesssim t_{\mathrm{sep}}$, leading to an intrinsic spatial distribution of pairs along the direction of the applied Bragg pulses. Averaging over the time of creation of the pairs results in the optimal offset between free propagation times before and after the applied mirror Bragg pulse. While this distinction may appear to only be a minor detail, it is important to emphasize that by using a pair of scattering halos in this manner our scheme does not require exhaustive calibration of the free propagation time. In contrast, the offset $t_{\mathrm{sep}}$ in Ref.~\cite{PhysRevA.91.052114} is derived under an assumption of a Gaussian density profile for the colliding BECs and would require experimental optimization.

Ultimately, the phase-sensitive correlations encoded by \(\mathcal{C}_{\textbf{L},\textbf{R}}\) can be used as part of a Bell inequality to demonstrate the inconsistency of some predictions of quantum mechanics with local hidden variable theories. First, we construct the quantum correlator, 
\begin{align}
    E(\phi_L, \phi_R) &= \frac{\mathcal{C}_{\textbf{p}, \textbf{q}'} + \mathcal{C}_{\textbf{p}', \textbf{q}}-\mathcal{C}_{\textbf{p}, \textbf{p}'}-\mathcal{C}_{\textbf{q}, \textbf{q}'}}{\mathcal{C}_{\textbf{p}, \textbf{q}'} + C_{\textbf{p}', \textbf{q}}+\mathcal{C}_{\textbf{p}, \textbf{p}'}+\mathcal{C}_{\textbf{q}, \textbf{q}'}} \label{eqn:E_def} \\
    &=\frac{h \alpha_x \alpha_y \beta_z}{16 \prod_{d} \lambda_d^{2} + h \alpha_x \alpha_y \beta_z} \cos \left( \phi_L + \phi_R\right) , \label{eqn:E_final}
\end{align}
which depends only on the global phase $\Phi=\phi_L+\phi_R$ imprinted during the interferometer.
From this, one obtains the CHSH-Bell parameter $S$ for the Clauser-Horne-Shimony-Holt (CHSH) version of the Bell inequality \cite{PhysRevLett.64.2495,PhysRevLett.23.880}, where
\begin{equation}
S=|E(\phi_L, \phi_R)-E(\phi_L, \phi_R')+E(\phi_L', \phi_R)+E(\phi_L', \phi_R')|, \label{eqn:CHSH}
\end{equation}
which is bounded by $S \leq 2$  for any local hidden variable theory. For $E(\phi_L,\phi_R)$, as given by Eq.~\eqref{eqn:E_final}, we find that it is possible to violate this inequality for specific choices of phase settings \cite{PhysRevLett.64.2495,PhysRevA.91.052114} 
and instead we are limited by $S \leq 2\sqrt{2}$. As our interferometer depends on the global phase $\phi_L + \phi_R$, as opposed to the phase difference as in Refs.~\cite{PhysRevLett.64.2495,PhysRevA.91.052114}, we note that the optimal phase settings are different to the typical choice by a sign, $(\phi_L,\phi^{\prime}_L,\phi_R,\phi^{\prime}_R) = (0,\pi/2,-\pi/4,-3\pi/4)$~\footnote{For complete clarity, we note that the difference in phase sensitivity in our scheme to, e.g., Refs.~\cite{PhysRevLett.64.2495,PhysRevA.91.052114}, is due to an effective switching of which arm of the interferometer one of the phase shifts is imprinted.}. The saturation of the quantum bound, $S = 2\sqrt{2}$, occurs when we take $\lambda_{x,y,z} \to 0$ and $h \to \infty$, which corresponds to the limit where the twin-atoms of the scattering halo can be approximately mapped to the prototypical Bell state [Eq.~\eqref{eqn:BellstateHV}]. 
The conditions for this to occur typically correspond to a very low population of the scattering halo and large correlation length. Equation \eqref{eqn:E_final} is the main result of our analytic model, and the dependence on \(\Phi\) and \(\lambda_{x,y,z}\)  is compared to experimental data in the following section (see Figs.~\ref{fig:interference} and \ref{fig:E_lambda}, respectively).

\begin{table}[t]
    \centering
    \caption{Experimental parameters used in our interferometric protocol. The BEC number refers to the average number of atoms in the initial source BEC, before momentum splitting. The times $t_2$ and $t_3$ are qutoed with \(t_0=0\).}
    \begin{tabular}{ll}
    \toprule
    \toprule
        \(\left(\omega_x, \omega_y, \omega_z\right) /2 \pi\) & \( \left(49.607(3), 195.414(7), 201.21(7)\right) \) Hz \cite{Henson:22} \\
        \(\left(\sigma_{g,x}, \sigma_{g,y}, \sigma_{g,z}\right)\) & \(\left(26(3),4.2(3),4.1(3)\right)\) \(\mu\)m \\
        \(\left(\sigma_{x}, \sigma_{y}, \sigma_{z}\right)\) &  \(\left(3(1),15(4),8(3)\right)\) mm/s\\
        BEC Number & \(14(4)\, \times \, 10^4\)\\
        \(\bar{N}\) & \(0.15\) \\
        \(|\textbf{k}_0|\) & \(4.102\) \(\mu\)m\(^{-1}\) \\
        \(t_2\) & \(240\) \(\mu\)s\\
        \(t_3\) & \(480\) \(\mu\)s\\
        \(t_{\mathrm{sep}}\) & \(63(2)\) \(\mu\)s \\
    \bottomrule
    \bottomrule
    \end{tabular}
    \label{tab:params}
\end{table}

\section{Results}
\label{sec:results}

\begin{figure}[tbp]
    \centering
    \sidesubfloat[]{\includegraphics[width=0.85\textwidth]{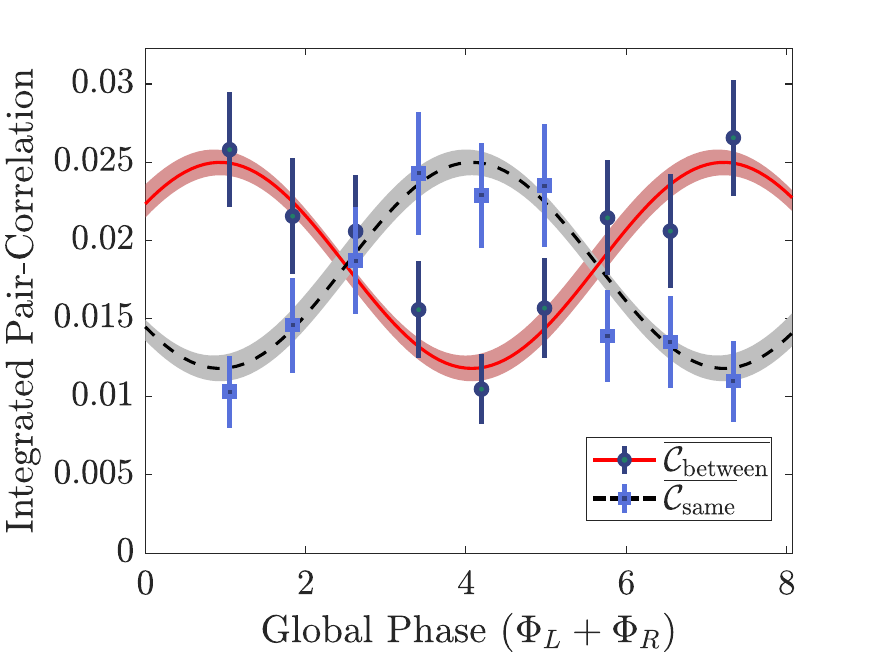}\label{fig:a}}\\[-1pt]
    \sidesubfloat[]{\includegraphics[width=0.85\textwidth]{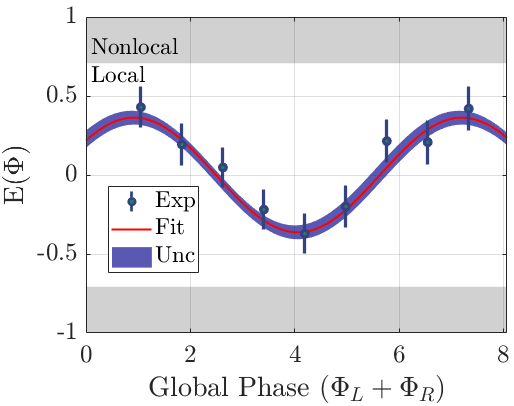}\label{fig:b}}
    \caption{Spatially separated momentum interference. (a) Phase dependence of average integrated pair correlation function for relevant port combinations (\(C_{\mathrm{same}}\) and \(C_{\mathrm{between}}\)). Error bars are generated from bootstrapping with replacement \cite{bootstrap}. A sinusoidal model of the form of Eq.~\eqref{eqn:corr_final} is fit to the entire data set with parameters \(\lambda = 0.6\), \(h=1.48\) and \(\bar{N}=0.15\), with shaded regions indicating fit uncertainty. The back-to-back correlation within the same halo \(C_{\mathrm{same}}\) being \(\pi\) out of phase relative to the correlations between halos \(C_{\mathrm{between}}\) is strong evidence of spatially separated interference, and hence entanglement. (b) Quantum correlator \(E(\Phi)\) versus global phase with sinusoidal fit with amplitude \(0.36(4)\) and phase offset \(-0.9(1)\). Again, error bars are produced using bootstrapping with replacement. This curve would produce a maximal possible Bell parameter of \(S=1.1(1)\), if we assume the independent phases follow the global phase \cite{CHSH_BELL_NOTE}. The correlation amplitude required to demonstrate Bell inequality violation is indicated by the gray shade region.
    }
    \label{fig:interference}
\end{figure}
\begin{figure}[t]
    \centering
    \includegraphics[width=0.8\textwidth]{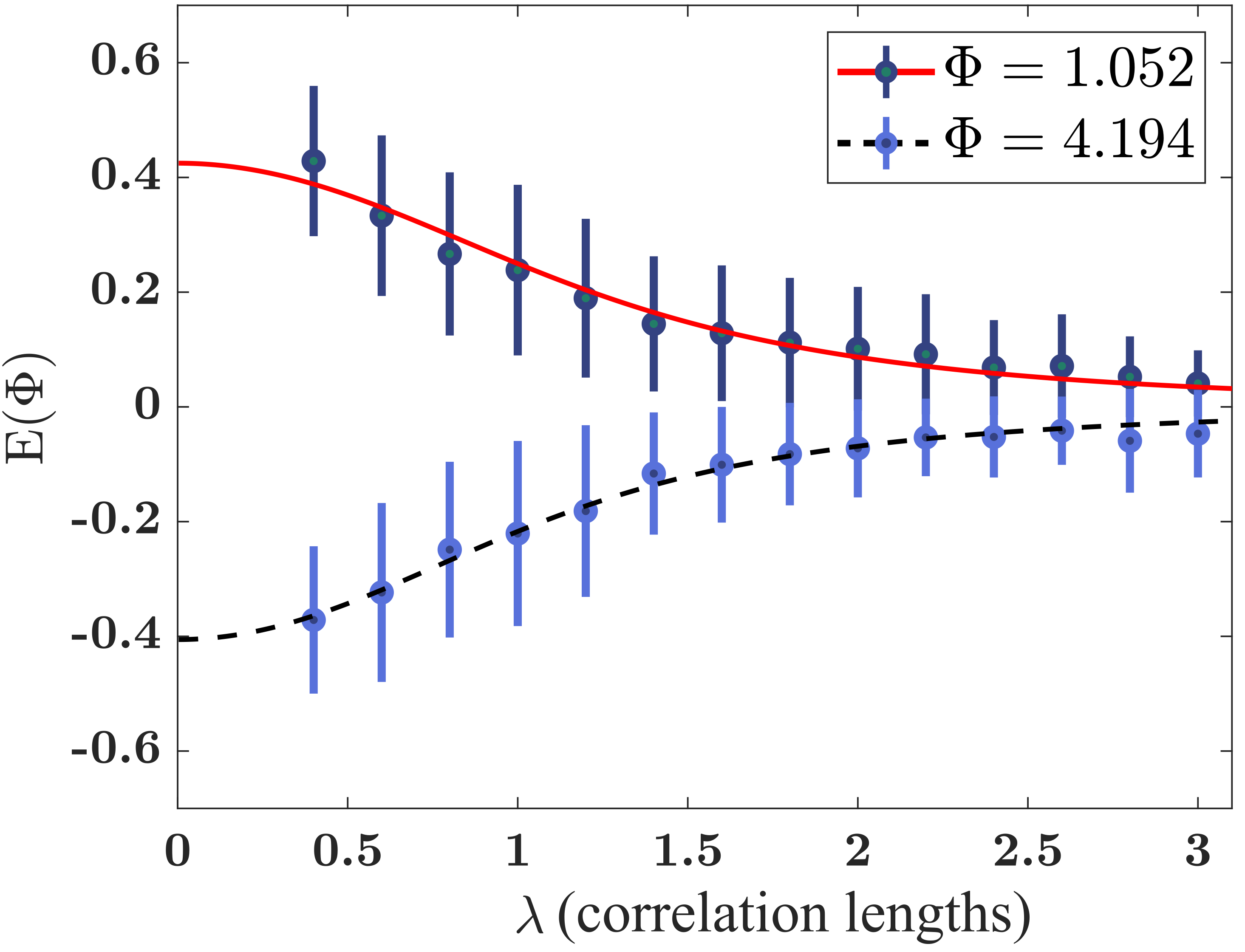}
    \caption{The quantum correlator \(E\), for global phases \(\Phi=1.052\) and \(\Phi=4.194\), versus normalised integration bin size, where for all dimensions \(d=x,y,z\) we have set \(\lambda_d=\lambda\). Note that \(\lambda\) is equivalent to expressing the integration bin size in terms of number of correlation lengths. Error bars on the experimental data points are produced by bootstraping with replacement \cite{bootstrap}. Solid lines represent fits of the form of Eq.~\eqref{eqn:E_final} with parameters \(h=1.5(1)\) and \(h=1.4(1)\) for \(\Phi=1.052\) and \(\Phi=4.194\) respectively, with \(t_3/2 = t_2\) and \(\{\sigma_{x}, \sigma_{y}, \sigma_{z}\}=\)\(\{3(1),15(4),8(3)\}\) mm/s for both data sets. The \(h\) parameters correspond to optimal values of \(E(1.052)\vert_{\lambda_{x,y,z}\rightarrow 0}=0.42(5)\) and \(E(4.194)\vert_{\lambda_{x,y,z}\rightarrow 0}=-0.40(5)\), respectively. 
    }
    \label{fig:E_lambda}




\end{figure}

In order to demonstrate our protocol's capabilities for spatially separated two particle interference between momentum states, we employ the Rarity-Tapster scheme as shown in Figs.~\ref{fig:concept_pos_sapce} and \ref{fig:concept}, with experimental parameters as listed in Tab.~\ref{tab:params}. We emphasize that at the time of the beam-splitter [\(t_3=480 \, \mu\)s, step (iv)] the atoms are separated by \(62.4~\mu\)m, which corresponds to a separation of \(\sim 4\) spatial correlation lengths (along the axis of separation).

We first investigate the integrated pair-correlation function $\mathcal{C}_{\mathrm{L},\mathrm{R}}$, as defined in Eq.~\eqref{eqn:int_pair_corr}. The requirement of maintaining a low average mode occupancy to achieve large correlation amplitudes, in combination with the overall low detection efficiency available, makes maximizing the rate of data aquisition extremely important. 
To this end, we improve our data rate by averaging over an ensemble of momentum modes that are sufficiently close to the equatorial planes of the upper and lower scattering halos, that simultaneously realize a set of independent Rarity-Tapster interferometers within each experimental trial.  
This is motivated by the fact that there are multiple sets of four momentum modes within the two halos which satisfy the resonance conditions of Eq.~\eqref{eqn:int_conditions}.
Upon doing such averaging, we account for the fact that our Bragg pulses are imperfect -- i.e., they have a finite width in momentum space over which they realize precise mirror and beamsplitter operations. 
To satisfy this constraint we select pairs within an azimuthal angle of \(20^\circ\) from the equator of either halo, which in turn corresponds to a velocity spread of \(\sim 23\)~mm/s along the \(z\)-axis relative to each equatorial plane.
Mathematically such averaging can be expressed as,
\begin{multline}
    \overline{\mathcal{C}_{\mathrm{same}}} = \frac{1}{2\mathcal{V}^{\mathrm{up}}_{\mathrm{eq}}}\int_{\mathbf{p} \in \mathcal{V}^{\mathrm{up}}_{\mathrm{eq}}} d \textbf{p} \, \, \mathcal{C}_{\textbf{p},-\textbf{p} + 2\textbf{k}_0} \\ + \frac{1}{2\mathcal{V}^{\mathrm{low}}_{\mathrm{eq}}}\int_{\mathbf{q} \in \mathcal{V}^{\mathrm{low}}_{\mathrm{eq}}} d \textbf{q} \, \, \mathcal{C}_{\textbf{q},-\textbf{q} - 2\textbf{k}_0} ,
\end{multline}
and
\begin{multline}
    \overline{\mathcal{C}_{\mathrm{between}}} = \frac{1}{2\mathcal{V}^{\mathrm{up}}_{\mathrm{eq}}}\int_{\mathbf{p} \in \mathcal{V}^{\mathrm{up}}_{\mathrm{eq}}} d \textbf{p} \, \, \mathcal{C}_{\textbf{p},-\textbf{p}}  \\  + \frac{1}{2\mathcal{V}^{\mathrm{low}}_{\mathrm{eq}}}\int_{\mathbf{q} \in \mathcal{V}^{\mathrm{low}}_{\mathrm{eq}}} d \textbf{q} \, \, \mathcal{C}_{\textbf{q},-\textbf{q}} ,
\end{multline}
where $\mathcal{V}^{\mathrm{up}}_{\mathrm{eq}}$ and $\mathcal{V}^{\mathrm{low}}_{\mathrm{eq}}$ are the relevant averaging volumes of the equatorial planes of the upper and lower halos, respectively. 
The two definitions correspond to the correlations of approximately diametrically opposing modes within a single halo (\(\overline{\mathcal{C}_{\mathrm{same}}}\)) and between the two halos (\(\overline{\mathcal{C}_{\mathrm{between}}}\)).

We present the average integrated pair correlation function for both port combinations as a function of global phase in Fig.~\ref{fig:interference}(a). We find strong sinusoidal dependences, reflected by a good fit to Eq.~\eqref{eqn:corr_final} for both \(\overline{\mathcal{C}_{\mathrm{same}}}\) and \(\overline{\mathcal{C}_{\mathrm{between}}}\) with the amplitude of the oscillatory correlation taken to be a free parameter, with an coefficient of determination (\(R^2\)) of \(0.88\). Moreover, we clearly observe the expected \(\pi\) phase offset between the two data sets. The amplitude of the theory fit is \(0.0066(6)\), while the interferometric visibility of the \(\overline{\mathcal{C}_{\mathrm{between}}}\) data is \(0.40(15)\) and the \(\overline{\mathcal{C}_{\mathrm{same}}}\) data is \(0.44(12)\), giving an average interferometric visibility of \(V=0.42(9)\). 

We now turn to extracting the magnitude of the quantum correlator \(E\), Eq.~\eqref{eqn:E_def}, from the experimental data, which is the key quantity for evaluating the value of the CHSH-Bell parameter \(S\). The quantum correlator \(E\), plotted in Fig.~\ref{fig:interference}(b) as a function of the global phase, has an amplitude of \(0.36(4)\), corresponding to a confidence level of \(9\sigma\) above zero. However, this does not surpass the value \(|E| = 1/\sqrt{2}\), which would be required for a violation of the CHSH-Bell inequality \cite{PhysRevLett.23.880}. Instead our result  is indistinguishable from the predictions of a local hidden variable theory, satisfying \(|E| < 1/\sqrt{2}\) and thus \(S\leq 2\). We note that our extracted $E$ features a phase offset (similarly observable in \(\overline{\mathcal{C}_{\mathrm{same}}}\) and \(\overline{\mathcal{C}_{\mathrm{between}}}\)). This is expected due to a gravitational phase shift (See Appendix~\ref{sec:grav_phase}, particularly Eq.~\eqref{eqn:phase_dif}), although the experimental value of \(-0.9(1)\) is two standard deviations larger in magnitude than the theoretical prediction \(-0.5(2)\).

Beyond demonstrating the expected sensitivity of the experimentally obtained correlations to the imprinted global phase, we can also compare to the predicted dependence on, e.g., the correlation width and integration volume, from the the analytic model. 
In Fig.~\ref{fig:E_lambda} we compare the experimental dependence of the quantum correlator \(E\) on integration bin size \(\lambda\) to the functional form predicted by Eq.~\eqref{eqn:E_final}. For the theoretical prediction we use the known values of helium mass, $\Phi$, $k_0$, $t_2$ and $t_3$ of the applied Bragg pulses, and also the correlation widths $\sigma_{x,y,z}$ that are obtained from the initial scattering halos \cite{PhysRevLett.118.240402}. This leaves the correlation height $h$ as a free parameter.
We find good agreement  between the analytical model and empirical data (\(R^2=0.972\)). This gives strong evidence that our model and the underlying assumption that the initial pair correlations, $G^{(2)}(\mathbf{k},\mathbf{k}',t_1)$, are well described by the Gaussian approximation, Eq.~\eqref{eqn:gaussian_g2}. However, the fit implies we have a small correlation height above the background, about \(h=1.4(1)\). This is despite our mode occupancy indicating a much stronger correlation height: For the mode occupancy \(\bar{N} \sim 0.15\) used in this work we expect \(h \sim 7.6\), and we have previously confirmed that this relationship holds in scattering halos with no applied Bragg Pulses \cite{PhysRevLett.118.240402,Shin2019}.
\revcom{The mode occupancy $\bar{N}$ is measured experimentally by dividing the average number of particles in each scattering halo by the number of modes in the experimental volume (set by the measured momentum correlation widths, given in Tab.~\ref{tab:params} for values used in this work), see supplementary material of Refs.~\cite{PhysRevLett.118.240402} and \cite{Perrin2008} for further detail.}

This implies that there are factors degrading our interfometetric signal. There are a number of known issues beyond finite correlation width, detection resolution, and free propagation time that are not included in our model, such as Bragg pulse efficiency and unequal mode occupancy between the halos. We quantify the Bragg pulse peak efficiency (which we find to be \(0.984\)) and momentum distribution (see Appendix Fig.~\ref{fig:halo}) in order to measure the possible effects of imperfect Bragg pulses on the correlation height $h$. Initial estimates indicate that the magnitude of these effects are small, and hence they alone are not enough to explain the decrease in signal. 
\revcom{As explained in Sec.~\ref{sec:Theory} our system is a probabilistic generator of particle entangled Bell states. This primarily has the effect of worsening our statistics and a reduction in signal height due to the inclusion of states with higher occupation, which is encapsulated by the relation \(h\approx1/\bar{N}\) and is independent of detector efficiency (approximately \(8\%\), which we estimate using relative number squeezing \cite{PhysRevLett.105.190402}). However, it is also possible to include false positives due to dark counts, which will further degrade our interfometetric signal. Our measured estimates of the dark count rates are very small (less than 0.01 counts per millimeter squared per second corresponding to an average of less than \(6\times10^{-5}\) dark counts per integration volume for \(\lambda=0.5\), i.e. a single mode volume) and a false positive would require a simultaneous dark count to appear in at least two of the relevant momentum modes. Hence this also should not be a significant effect compared to our mode occupancy of \(\sim 0.15\). However, it cannot be conclusively ruled out. Note that the effect of dark counts is also independent of detector efficiency. }
Some further possible sources are gravitational effects over the finite wavepacket size of the scattered pairs, and mean-field effects due to the source condensates \cite{PhysRevLett.104.150402,Deuar2011,PhysRevA.87.061603}, both of which we have yet to properly quantify. Incorporating these effects into the model and overcoming them experimentally will be the focus of our future work on this system.




\section{Conclusion}
\label{sec:con}
We have demonstrated a matter wave Rarity-Tapster interferometer and thus an experimentally viable method for generating interference between the momentum states of spatially separated helium atoms. The method is underpinned by entanglement between twin-atoms generated by atomic four-wave mixing and exploits the geometry of scattering halos formed from multiple colliding BECs.

The primary advantage of the method we present is that it allows for efficient optimization and alignment of the interferometer due to the ability to selectively generate only a single halo (equivalently a single set of modes \((\textbf{p},\textbf{p}')\) or \((\textbf{q},\textbf{q}')\) as described above) for calibration purposes, and requires only a single set of Bragg beams, making it simple to implement experimentally and more stable and robust against alignment drifts. 
Due to the specific geometry, our interferometer is sensitive to \emph{global} phase, i.e., the \emph{sum} of phases $\phi_L$ and $\phi_R$ in the two arms of the interferometer, in contrast to a traditional Rarity-Tapster interferometer which is sensitive to the respective \emph{relative} phase. This enables us to readily test its interference capabilities without the technically demanding implementation of a spatially dependent phase, although independent control of the applied phases $\phi_L$ and $\phi_R$, as in Eq.~\eqref{eqn:CHSH}, is needed to demonstrate a formal violation of the CHSH-Bell inequality.

As a proof of concept we demonstrate phase-sensitive integrated pair correlation functions with an average visibility of \(V=0.42(9)\) and quantum correlator amplitude of \(E_0 = 0.36(4)\), underpinned by efficient Bragg pulses (with peak transmission efficiency of \(0.984\)) and characterization of the momentum distribution of the scattering halos (see Appendix Fig.~\ref{fig:halo}).

This method could also be extended to higher order interference schemes, for either more atoms or more modes, as well as to prototype more complex Bragg pulses. We hope that in future this work will lead to the demonstration of a full violation of a Bell inequality, and more generally be leveraged as the basis for momentum-space interferometry experiments with massive particles.

\begin{acknowledgments}

This  work  was  supported  through Australian  Research  Council  (ARC)  Discovery  Project Grant
DP190103021.
K.F.T. was  supported  by  Australian  Government  Research Training  Program  (RTP)  scholarships.
\end{acknowledgments}

\section*{Author contributions}

The experiments were carried out by KFT, YW and BMH, under the supervision of SSH and AGT. Data analysis was performed by KFT, under the supervision of SSH and AGT. The theoretical model was developed by RJLS and KVK. The modified experimental setup was developed by BMH and KFT. All authors were involved with preparation of the manuscript, with the first draft written by KFT, RJLS, and KVK.

\vspace{0.5cm}
\textbf{Data Availability Statement} The authors confirm that the data supporting the findings of this study are available within the article and its supplementary materials.

\begin{appendix}

\section{Derivation of the mapping between input and output states of our interferometer}
\label{sec:mapping}
In this section we will follow the same procedure as Appendix Bragg diffraction model of Ref.~\cite{PhysRevLett.119.173202}. The coupling Hamiltonian of our Bragg pulse can be written as \cite{meystre2001atom,PhysRevLett.119.173202},
\begin{align}
    \hat{H} &= \frac{\hbar \Omega}{2} \begin{pmatrix}
0 & e^{i\phi} \\
e^{-i\phi}  & 0
\end{pmatrix},
\end{align}
where we have used the basis \(\{\hat{a}_{\textbf{k}} ,\hat{a}_{\textbf{k}+2\textbf{k}_0}\}\), \(\Omega/2\pi\) is the two-photon Rabi frequency, and \(\phi\) is the phase of the Bragg lattice, which is operationally set by the phase difference of the lasers that form the lattice. The evolution operator is defined as,
\begin{align}
    \hat{U}(t,\phi) &\!=\! e^{-i\hat{H} t/\hbar} \!=\! \begin{pmatrix}
\cos\frac{\Omega t}{2} & -ie^{-i\phi} \sin \frac{\Omega t}{2}  \\
-ie^{i\phi} \sin \frac{\Omega t}{2}   & \!\cos\frac{\Omega t}{2} 
\end{pmatrix}.
\end{align}
The dynamics of our the Rarity-Tapster setup can then be modeled as the application of a \(\pi\)-pulse and \(\pi/2\)-pulse, i.e. \(\hat{U}( \pi/2\Omega,\phi_{\pi/2})\hat{U}(\pi/\Omega,\phi_\pi)=\hat{A}\), which can be written as
\begin{align}
   \hat{A} &= -\frac{1}{\sqrt{2}}\begin{pmatrix}
e^{-i(\phi_{\pi/2}-\phi_{\pi})} & ie^{-i\phi_{\pi}} \\
ie^{i\phi_{\pi}}  & e^{i(\phi_{\pi/2}-\phi_{\pi})}
\end{pmatrix},
\end{align}
where \(\phi_{\pi/2}\) 
is the phase of the \(\pi/2\) (beamsplitter) Bragg pulse and \(\phi_{\pi}\) is the phase of the \(\pi\) (mirror) pulse. Note, this model does not currently contain the effect of realistic Bragg pulses, instead assuming they are instantaneous perfect linear transformations and contains no information on the timings of pulses, as at this point we are assuming the modes have an infinite spatial extent. In order to obtain the input modes as functions of the output modes we invert the operator \(\hat{A}\) matrix,
\begin{align}
    \hat{A}^{-1}&= -\frac{1}{\sqrt{2}}\begin{pmatrix}
e^{i(\phi_{\pi/2}-\phi_{\pi})} & -ie^{-i\phi_{\pi}} \\
-ie^{i\phi_{\pi}}  & e^{-i(\phi_{\pi/2}-\phi_{\pi})}
\end{pmatrix}.
\end{align}
As the phase of the mirror does not affect the dynamics we can choose it for convenience to be \(\phi_{\pi}=\pi/2\).
Using \(A^{-1}\) we can write our input modes \(\hat{b}\) in terms of our output modes \(\hat{a}\), taking into account the phase of the beamsplitter for the left and right arms, as follows,
\begin{align}
\begin{pmatrix}
\hat{b}_{\textbf{p}} \\
\hat{b}_{\textbf{q}}
\end{pmatrix} &= \frac{1}{\sqrt{2}}\begin{pmatrix}
ie^{i\phi_R} & 1 \\
-1  & -ie^{-i\phi_R}
\end{pmatrix} \begin{pmatrix}
\hat{a}_{\textbf{p}} \\
\hat{a}_{\textbf{q}}
\end{pmatrix},\label{eqn:first_map_app}\\
\begin{pmatrix}
\hat{b}_{\textbf{p}'} \\
\hat{b}_{\textbf{q}'}
\end{pmatrix} &= \frac{1}{\sqrt{2}}\begin{pmatrix}
ie^{i\phi_L} & 1 \\
-1  & -ie^{-i\phi_L}
\end{pmatrix} \begin{pmatrix}
\hat{a}_{\textbf{p}'} \\
\hat{a}_{\textbf{q}'}
\end{pmatrix}.\label{eqn:second_map_app}
\end{align}

As an exemplar we let the initial state \(\ket{\Psi}\) be the prototypical Bell state (see Eq.~\eqref{eqn:proto_bell}). Using Eqs.~\eqref{eqn:first_map_app} and \eqref{eqn:second_map_app} the output state after propagation through the interferometer  can be written as
\begin{align}\label{eqn:BellStateOut}
    \ket{\Psi}_{\mathrm{out}} =  \frac{1}{2\sqrt{2}} \Big[\left(1-e^{i(\phi_L+\phi_R)}\right) &\ket{1}_{\mathbf{p}}\ket{1}_{\mathbf{p'}}\ket{0}_{\mathbf{q}}\ket{0}_{\mathbf{q}'} \nonumber\\+i(e^{i\phi_L}+ e^{-i\phi_R})  &\ket{0}_{\mathbf{p}}\ket{1}_{\mathbf{p'}}\ket{1}_{\mathbf{q}}\ket{0}_{\mathbf{q}'} \nonumber\\  +i(e^{i\phi_R} + e^{-i\phi_L}) &\ket{1}_{\mathbf{p}}\ket{0}_{\mathbf{p'}}\ket{0}_{\mathbf{q}}\ket{1}_{\mathbf{q}'} \nonumber\\
     + \left(1- e^{-i(\phi_L+\phi_R)}\right) &\ket{0}_{\mathbf{p}}\ket{0}_{\mathbf{p'}}\ket{1}_{\mathbf{q}}\ket{1}_{\mathbf{q}'} \Big].
\end{align}
The form of Eq.~\eqref{eqn:BellStateOut} can be readily compared to that of Eq.~(6) of Ref.~\cite{PhysRevLett.119.173202}, which shows the effect of an unmodified Rarity-Tapster scheme on a prototypical Bell state.

We observe that the measurement of correlations between multiple modes has a dependence on the applied phase shift. This is best illustrated by the pair-correlation function $C_{\mathbf{k},\mathbf{k}^{\prime}} = \langle \hat{a}^{\dagger}_{\mathbf{k}} \hat{a}^{\dagger}_{\mathbf{k}^{\prime}} \hat{a}_{\mathbf{k}^{\prime}} \hat{a}_{\mathbf{k}}\rangle$,
which corresponds to a correlation between joint-detection events at the outputs of the left and right arms of the interferometer (see Figs.~\ref{fig:concept_pos_sapce} and \ref{fig:concept}). For the prototypical Bell state, Eq.~\eqref{eqn:BellStateOut}, we obtain,
\begin{equation}
    \begin{gathered}\label{eqn:C_kkp_bell}
        C_{\mathbf{p},\mathbf{p}^{\prime}} = C_{\mathbf{q},\mathbf{q}^{\prime}} = \frac{1}{2}\sin^2\left( \frac{\phi_L + \phi_R}{2} \right) , \\
        C_{\mathbf{p},\mathbf{q}^{\prime}} = C_{\mathbf{p}^{\prime},\mathbf{q}} = \frac{1}{2}\cos^2\left( \frac{\phi_L + \phi_R}{2} \right) .
    \end{gathered}
\end{equation}
This result crucially indicates that interference between the scattered pairs can be tuned via the \emph{global} phase \(\Phi=\phi_L+\phi_R\) and accessed via pair-correlation functions. This is in contrast to the previous implementations of the Rarity-Tapster scheme which are dependent on the \emph{relative} phase \(\phi_L-\phi_R\), see for instance Eqs.~(7) and (8) of Ref.~\cite{PhysRevLett.119.173202}.

\section{Spatial overlap and phase diffusion in the integrated correlation function}
\label{sec:phase_diffusion}

The result for the two-point correlation function $G^{(2)}(\textbf{k},\textbf{k}',t_4)$, given by Eq.~(\ref{eqn:output_g2_full}) of the main text, 
can be derived by following the calculation reported in Appendix B of Ref.~\cite{PhysRevA.91.052114}. The computation assumes that: (i) the applied Bragg pulses can be treated as simple linear transformations of the coupled momentum modes, and (ii) the two-point momentum correlation functions after the initial BEC collisions can be captured by a simple Gaussian ansatz (see Eq.~\eqref{eqn:gaussian_g2} of the main text). In addition, the calculation of Ref.~\cite{PhysRevA.91.052114} identifies that an effective dephasing arises, captured by $\varphi(\textbf{k},\textbf{k}')$, which, as discussed in the main text, can be identified as encoding the distinguishability of atoms with different momenta as they traverse the matter wave interferometer. An approximate form for $\varphi(\textbf{k},\textbf{k}')$ can be obtained via a perturbative treatment of the BEC collision \cite{PhysRevA.78.053605}. By adapting this perturbative treatment to the collision geometry of our experiment, we obtain (following analogous steps to Ref.~\cite{PhysRevA.91.052114}),
\begin{widetext}
\begin{equation}
\varphi(\textbf{k},\textbf{k}') =
\begin{cases}
 -\frac{2\hbar k_0}{m}(k_z + k^{\prime}_z \pm 2k_0)\left(\frac{t_3}{2}-t_2\right), & \text{for} ~(\mathbf{k},\mathbf{k}') \approx (\mathbf{p},\mathbf{p}') ~ \text{or} ~ (\mathbf{q},\mathbf{q}'), \\[14pt]
 -\frac{2\hbar k_0}{m}(k_z + k^{\prime}_z)\left(\frac{t_3}{2}-t_2\right), & \text{for} ~(\mathbf{k},\mathbf{k}') \approx (\mathbf{q},\mathbf{p}') ~ \text{or} ~ (\mathbf{p},\mathbf{q}'), 
\end{cases}
\end{equation}
\end{widetext}
where the plus and minus sign of the first line is associated with the upper ($-$) and lower ($+$) halos, identically to $G^{(2)}(\textbf{k},\textbf{k}',t_4)$. This form of $\varphi(\textbf{k},\textbf{k}')$ is used in the main text for the evaluation of $C_{\mathbf{L},\mathbf{R}}$ (Eq.~\eqref{eqn:corr_final}).

\section{Phase drift due to gravity}
\label{sec:grav_phase}
In Fig.~\ref{fig:interference} we observe a phase offset in the experimentally obtained phase-sensitive correlations $E$ and $\mathcal{C}_{\mathbf{L},\mathbf{R}}$. The theoretical treatment of Sec.~\ref{sec:Theory} assumes an ideal system 
with no relative phase drifts due to external factors beyond the intended phase settings of the Bragg pulses. There will however be a drift in phase accrued due to the gravitational force exerted on the particles over their free-fall trajectories. To account for this phase drift we develop a semi-classical model, where the trajectories are treated classically.

Within our model, the total phase \(\phi^A\) accrued  over a trajectory \(A\) is \cite{Peters_2001}, 
\begin{align}
    \phi^A &= \mathcal{S}^A/\hbar,
\end{align}
where
\begin{align}
    \mathcal{S}^A &= \int_A  \mathcal{L}(t)\, dt
\end{align}
is the classical action for path \(A\) and we require \(\mathcal{S}^A \gg \hbar\).
For our free-fall problem, assuming a uniform gravitational field, the Lagrangian $\mathcal{L}$ is given by $\mathcal{L}(t)=m|\textbf{v}^A(t)|^2/2-mgz^A(t)$, where $\textbf{v}^A(t)$ and \(z^A(t)\) are the velocity vector and \(z\)-axis coordinate of the particle on path \(A\) and $g$ is the gravitational acceleration. Therefore, the phase accrued during the interferometric sequence will be
\begin{align}
    \phi^A &= \frac{\mathcal{S}^A}{\hbar} = \frac{m}{\hbar}  \int^{t_3}_{t_0}  \left[\frac{1}{2} |\textbf{v}^A(t)|^2 - g z^A(t)\right]\,dt, \label{eqn:phi}
\end{align}
where \(t_0\) and \(t_3\) are the times of our initial and final pulses, as discussed in the main text. The velocity and position can be parameterised using the classical ballistic equations, and we can hence evaluate Eq.~\eqref{eqn:phi} analytically to give the phase difference between the upper (\(U\)) and lower paths (\(L\)) (for both arms of the interferometer),
\begin{align}
    \Delta \phi &= \phi^U(v_{z}(t_0),z(t_0)) - \phi^L(v_{z}(t_0)+ \Delta v_z,z(t_0))\nonumber \\ 
    &=\frac{2 g m t_1^2 \Delta v_z}{\hbar} =  k_{0} g t_3^2, \label{eqn:phase_dif}
\end{align}
where \(\Delta v_z\) is the momentum imparted by the mirror pulse, \(k_0=m \Delta v_z/2\hbar\) is the wave-vector of the Bragg lattice, and \(v_z(t_0)\) and \(z(t_0)\) are the initial \(z\)-components of the velocity and position of the particles, which both cancel out from the final expression.  Note that we have assumed that the pairs are all generated at the same time and position, that the modes exactly match the Bragg condition, and we have set \(t_3=2t_2\) as the total interferometer sequence takes twice the separation time. It can be seen that this is the standard expression for gravitational phase difference used in atom interferometry accelerometers \cite{Peters_2001}. However, a key difference to the Mach-Zehnder scheme discussed in Ref.~\cite{Peters_2001} is that the phase drift is applied to both arms of the interferometer (i.e., it is a drift in both \(\phi_L\) and \(\phi_R\)) and thus the change in the global phase will be \(\Delta \Phi = 2 \Delta \phi\). As we take gravity to be a uniform field this phase drift will be the same for all atom pairs across the two scattering halos.

\section{Theoretical basis of Bragg pulses}
\label{sec:bragg}
\begin{figure}
    \centering
    \sidesubfloat[]{\includegraphics[width=0.9\textwidth]{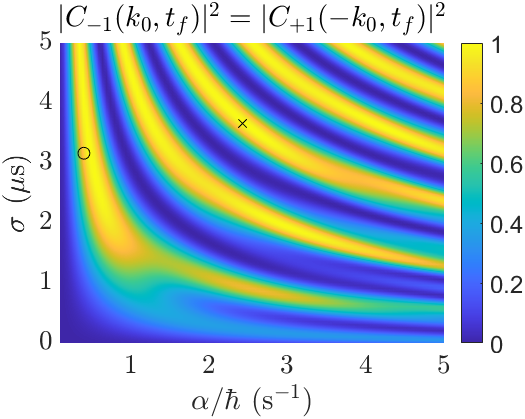}}\quad
    \sidesubfloat[]{\includegraphics[width=0.9\textwidth]{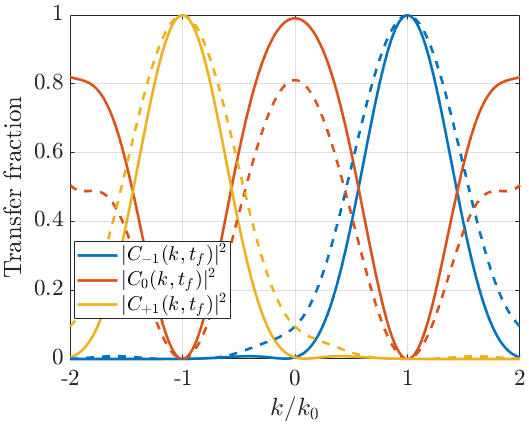}}
    \caption{Numerical Simulation of Bragg pulse. We numerically evaluate Eq.~\eqref{eqn:raman_nath}, with initial condition \(C_n(k,t=0)=\delta_{n,0}\), \(\delta=0\), \(\hbar k_0^2/2m=133.4\)~kHz and \(N=9\).
    (a) Transfer percentage at the halo equators \(k=\pm k_0\) against pulse parameters width \(\sigma\) and amplitude \(\alpha\), noting that the transfer is in the opposite direction for each equator, with parameters used for part (b) marked with an x and o. (b) Transfer versus momentum space for pulse \((\alpha,\sigma)\)=(0.405\(\hbar\) J, 3.162 \(\mu\)s) (solid line) and \((\alpha,\sigma)\)=(2.424\(\hbar\) J, 3.664 \(\mu\)s) (dashed line) showing how equal but opposite momenta are imparted to the halo equators at \(\pm k_0\).
    }
    \label{fig:Bragg_numeric}
\end{figure}




\begin{figure*}
    \centering
    \includegraphics[width=0.9\textwidth]{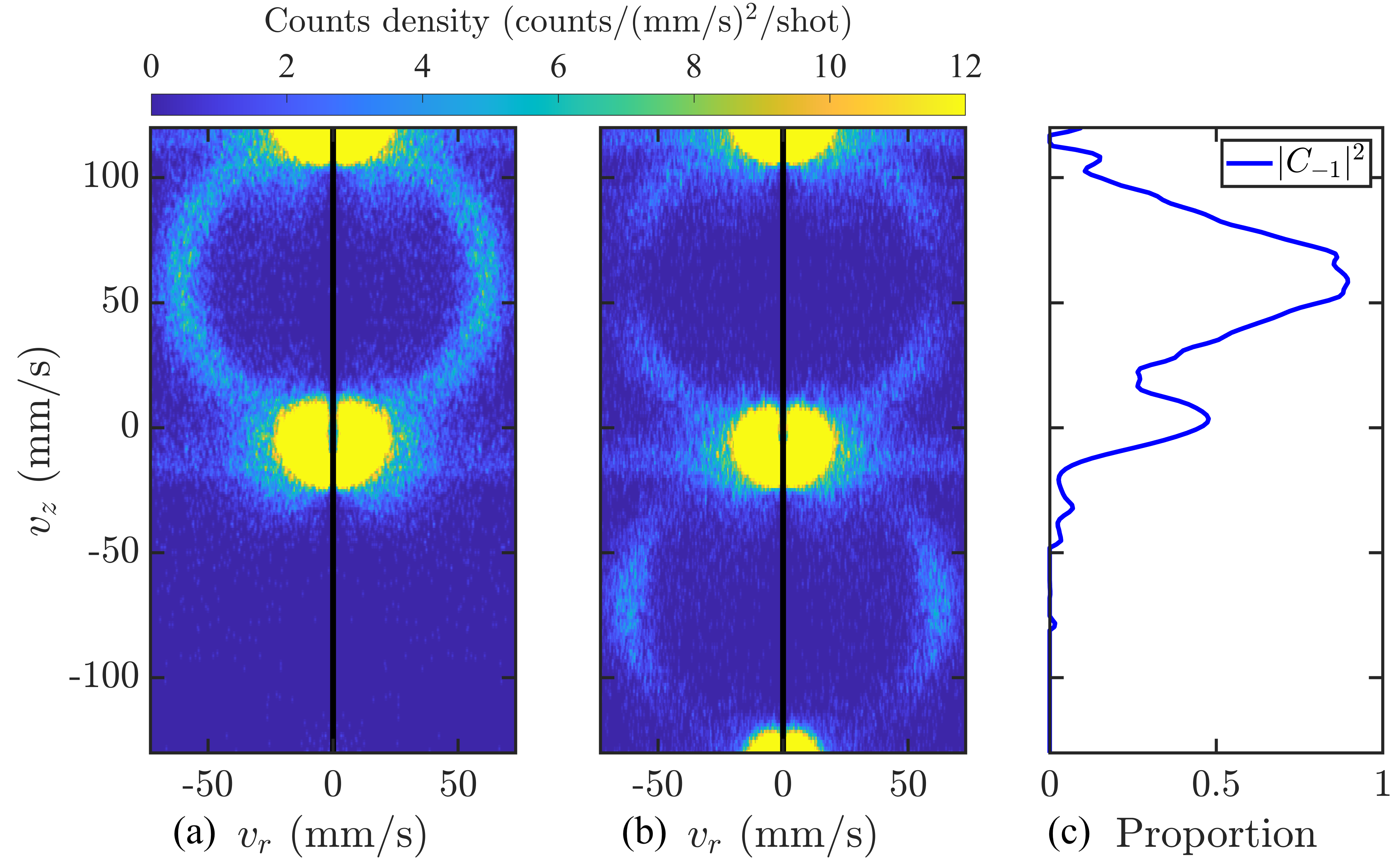}
    \caption{
    The effect of applying a single mirror pulse to the upper halo. (a) Velocity profile of the top halo; (b) velocity profile of same halo after a mirror pulse has been applied, allowing us to clearly see the effect of the pulse. The profiles are obtained by averaging over \(\sim 30\) experimental realisations. \revcom{The radial velocity \(v_r\) is an average of the values \(\sqrt{v_x^2+v_y^2}\). While strictly \(v_r\) cannot take negative values we mirror the profile about \(v_r=0\) to more clearly indicate the 3D cross section of the halo.} The black bar at \(v_r=0\) represents a break down of the histogram, as the detection bin volume approaches zero at this point. \revcom{Note that extra horizontal `spikes' emanating from the BECs are artifacts of the detection process, which occurs at high fluxes.} (c) The two profiles allow us to calculate the percentage of atoms transferred to each diffraction order for a particular momentum. Alternatively this can be thought of as the probability for a particle to be shifted a given number of Bragg vectors at that momentum. For simplicity we only show the momentum transfers \(|C_{-1}|^2\), as this is the distribution that can be most readily calculated from the presented data. The vertical axes for (a), (b), and (c) are all on the same scale.
    }
    \label{fig:halo}
\end{figure*}

We control the momentum states of our atoms via the use of Bragg pulses. These pulses are formed by counter-propagating laser beams, which interfere to form a periodic potential, and have a detuning such that they do not change the internal states of the atoms \cite{meystre2001atom}.

Let our counterpropagating lasers have wavevectors \(\textbf{k}_u\) and \(\textbf{k}_l\), and frequencies \(\omega_u\) and \(\omega_l\) respectively. To understand the effect of the Bragg pulses we model the interfering lasers as a standing wave potential \(\Omega(t) \cos (2 \textbf{k}_0 \cdot \textbf{z} + \delta t + \theta)\), where \(2 \textbf{k}_0 = \textbf{k}_u-\textbf{k}_l\) is the lattice vector, \(\delta = \omega_u -\omega_l\) is the frequency difference between the counter propagating beams, \(\Omega(t)\) is the amplitude modulation of the wave packet of the Bragg pulse, and \(\theta\) is the phase difference of the beams. The wave-vector $\mathbf{k}_0$ is aligned along the $z$ axis, with $|\mathbf{k}_0|\equiv k_0$. To proceed, we expand the wavefunction of the atoms in the empty lattice Bloch basis, \(\psi(\textbf{r},t)= \int d \textbf{k} \Sigma_n C_n(\textbf{k},t) e^{i(2n\textbf{k}_0+\textbf{k})\cdot \textbf{r}}\). If we ignore mean-field effects and approximate the lattice potential as infinite then the Schr\"odinger equation simplifies to the Raman-Nath equations \cite{PhysRevA.71.043602} for the expansion coefficients, 
\begin{align}
    i \frac{d}{dt} C_n(\textbf{k},t) &= \frac{\hbar}{2 m} (2n\textbf{k}_0 +\textbf{k})^2 C_n(\textbf{k},t) +\nonumber\\ 
    &\hspace{-0.3cm}\frac{\Omega(t)}{2} \left[e^{-i(\theta+\delta t)}C_{n-1}(\textbf{k},t) +e^{i(\theta+\delta t)} C_{n+1}(\textbf{k},t) \right]. \label{eqn:raman_nath}
\end{align}
From this we see that modes are only coupled to one another if they are separated by an integer number of lattice vectors (\(2\textbf{k}_0\)). In practice this is not strictly true due to various factors, such as mean-field interaction and the lattice having a finite spatial extent, however, Eq.~\eqref{eqn:raman_nath} is still a good approximation for realistic regimes and our experiment.

For our analysis, we will ignore the effect of the detuning \(\delta\) as it is equivalent to a reference frame shift and hence does not change any of the relevant dynamics. Thus, we can set \(\delta = 0\) without loss of generality, while experimentally we would tune \(\delta\) to couple the desired modes of the interferometer.

Let the initial mode population be \(C_n(\textbf{k},t=0)=\delta_{n,0}\), with \(|\textbf{k}|<\textbf{k}_0\), then we can consider \(|C_n(\textbf{k},t_f)|^2\) as the proportion of the initial mode population that is transferred by \(n\) lattice vectors after the Bragg pulse is applied, where \(t_f\) is some final time well after the Bragg pulse's time extent. While Eq.~\eqref{eqn:raman_nath} is an infinite set of coupled equations we can make numerical solutions tractable by truncating it to the \(M\) lowest-order equations, i.e., set \(C_n(k,t)=0\) for \(|n|>M\) and all \(t\). This approximation is valid for \(\Omega/2 \ll (2M)^2 \frac{\hbar k_0^2}{2 m}\) \cite{PhysRevA.71.043602}. Importantly, only the \(n=\pm 1\) modes are coupled for \(\Omega/2 \ll 4 \frac{\hbar k_0^2}{2 m}\), which is what we desire for our mirror and beamsplitter Bragg pulses.

We consider a Gaussian enveloped pulse \(\Omega(t) = \frac{\alpha}{\hbar} \exp\left[-(t-t_0)^2/(2\sigma^2)\right]\) and solve Eq.~\eqref{eqn:raman_nath} numerically with \(M = 9\) over the parameter space \((\sigma,\alpha)\) to give an insight into the dynamics of the Bragg pulse, see Fig.~\ref{fig:Bragg_numeric}(a). There are a number of possible parameter combinations which produce a peak diffraction efficiency of 1 or 0.5 (corresponding to a mirror or beam-splitter), however these pulses' distributions in momentum space can vary dramatically; see Fig.~\ref{fig:Bragg_numeric}(b) for comparison of two possible mirror pulse configurations. It is also determined that the optimal detuning (or centering in momentum space) of the pulses is about the halo equators (i.e., peak diffraction efficiencies at \(\pm k_0\)), as this represents the highest average transfer efficiency across the entirety of the halo. Alternatively, we can think of this as capturing the largest portion of atoms in our Bragg pulses.




While this model gives us a prediction of the optimal parameter regime for our Bragg pulses, practical implementation can cause a number of divergences from the idealized model. We hence empirically optimized our Bragg pulses parameters, such as intensity and width, for the desired pulse characteristics, guided by the results from Eq.~\eqref{eqn:raman_nath} and Fig.~\ref{fig:Bragg_numeric}. 



In Fig.~\ref{fig:halo} we demonstrate how the single halo generation can be used to determine the momentum transfer distribution of the Bragg pulse, allowing us to optimise our pulses for various purposes. It can be readily seen that the mirror pulse affects a finite width in momentum space as discussed above, indicating how it is impossible to obtain perfect mirror and beam-splitter pulses. Due to this, we only use a section of the halo about the equator, which we tune the Bragg pulses to be centered on, to calculate our correlation functions. This masking is defined by \(\theta_{tol}\), which is the maximum angular deviation a particle can have from the equator.

The Bragg pulses can add a phase difference between momentum states, which depends on the phase difference of the lasers \(\theta\). This allows us to easily add a global phase to the interferometer, as the phase difference in the beam splitter pulse will map, using the same notation as above, to \(\phi_L=\theta\) and \(\phi_R=\theta\), and hence \(\Phi = 2\theta\). Furthermore, as our experiment is sensitive to global phase our Bragg lasers are sufficient to achieve a spatially separated interference pattern.

Finally, we wish to highlight that only a single set of Bragg lasers is required for this entire protocol. This is of interest as the possible geometries of an interferometer which only uses a single set of Bragg lasers is severely constrained.

\section{Analysis details}

The phase \(\Phi\) directly corresponds to twice the phase difference between the two channels of the Keysight 33600A Series waveform generator that controls the respective AOMs for the upper and lower beams, which in turn controls their intensities. In Fig.~\ref{fig:interference} of the main text, we present data for nine phase values \(\Phi \in \{1.053,1.838,2.624,3.409,4.194,4.980,5.765,6.551,7.336\}\); the data were generated using \(\sim 2900\) experimental runs for each phase.


To estimate the uncertainty in the experimentally obtained data, we use a bootstrapping technique. This consists of repeating the analysis procedure, for instance calculating the correlation functions, on subsets of the total data set, where the subsets can be drawn with replacement, i.e. a particular data point can appear more than once in a given subsample. The estimated uncertainty is then the variance between the full set of estimated values, weighted for sample size. This technique is based upon using the empirical distribution function as an approximation for the true distribution function.

\end{appendix}

\bibliography{refs.bib}

\begin{thebibliography}{72}
\bibitem [{CHS()}]{CHSH_BELL_NOTE}%
  \BibitemOpen
  \href@noop {} {}\bibinfo {note} {Measuring the full CHSH Bell parameter, as
  defined in Eq.~\ref{eqn:CHSH}, would require independent control of the
  phases imprinted on the two arms of the interferometer, which we have not
  currently implemented.}\BibitemShut {Stop}%
\end{thebibliography}
\end{document}